\documentclass[12pt, oneside]{article}   	

\usepackage{geometry}                		 
\usepackage{graphicx}					
\usepackage{amssymb}
\usepackage{amsmath}
\usepackage{amsthm}
\usepackage{array}
\usepackage[singlelinecheck=false]{caption}
\usepackage[usenames,dvipsnames,svgnames,table]{xcolor}
\usepackage{dsfont}
\graphicspath{ {Figures/} }
\usepackage{listings}
\usepackage{framed}
\usepackage{multirow} 
\usepackage{arydshln}
\usepackage{subcaption}
\usepackage{bbm}
\usepackage{makecell}
\usepackage{eufrak}
\usepackage{algorithm} 
\usepackage{algpseudocode}
\usepackage{hyperref}

\usepackage{etoolbox}

\patchcmd{\thebibliography}{%
  \section*{\refname}%
}{%
  \subsection*{\refname}%
}{}{}

\apptocmd{\thebibliography}{%
  \setlength{\itemsep}{0pt}%
  \setlength{\parskip}{0pt}%
}{}{}

\theoremstyle{remark}

\theoremstyle{definition}

\newcommand{\p}{\partial}
\newcommand{\ul}{\underline}

\def\m1{\mathbbm{1}}

\usepackage{tabularx}
\makeatletter
\def\hlinewd#1{%
\noalign{\ifnum0=`}\fi\hrule \@height #1 %
\futurelet\reserved@a\@xhline}
\makeatother

\numberwithin{equation}{section}

\begin{document}
\lstset{language=Matlab} 

%\title{Radial Basis Function Methods for Partial Differential Equations}
%\title{Recent Advances on Radial Basis Function Methods in Application to Partial Differential Equations}
\title{Pricing Derivatives \\ under Multiple Stochastic Factors  \\ by Localized Radial Basis Function Methods}

\author{Slobodan Milovanovi\'{c} \\ Uppsala University \\ \href{mailto:slobodan.milovanovic@it.uu.se}{slobodan.milovanovic@it.uu.se}
\and Victor Shcherbakov\\ Uppsala University \\ \href{mailto:victor.shcherbakov@it.uu.se}{victor.shcherbakov@it.uu.se} }

\date{}

\maketitle

\begin{abstract}
We propose two localized Radial Basis Function (RBF) methods, the Radial Basis Function Partition of Unity method (RBF--PUM) and the Radial Basis Function generated Finite Differences method (RBF--FD), for solving financial derivative pricing problems arising from market models with multiple stochastic factors. We demonstrate the useful features of the proposed methods, such as high accuracy, sparsity of the differentiation matrices, mesh-free nature and multi-dimensional extendability, and show how to apply these methods for solving time-dependent higher-dimensional PDEs in finance. We test these methods on several problems that incorporate stochastic asset, volatility, and interest rate dynamics by conducting numerical experiments. The results illustrate the capability of both methods to solve the problems to a sufficient accuracy within reasonable time. Both methods exhibit similar orders of convergence, which can be further improved by a more elaborate choice of the method parameters. Finally, we discuss the parallelization potentials of the proposed methods and report the speedup on the example of RBF--FD.
\end{abstract}

%%%%%%%%%% --- INTRODUCTION --- %%%%%%%%%%%%
\section{Introduction}\label{sec:Introduction}

Pricing derivatives and calibration of financial models are computationally very intensive tasks, which require a certain precision. Therefore, models that should be used are the ones which allow for the most accurate representation of market features, such as volatility smiles or skews and fat tails of the return distributions. Often, such market models involve multiple stochastic factors, e.g., stochastic volatility and interest rate. Although, each additional stochastic factor gives rise to an extra dimension. Thus, having several stochastic factors leads to multi-dimensional problems. However, thanks to the development of modern computational hardware and numerical methods, multi-factor asset pricing models rapidly gain in popularity. In this paper we suggest a numerical approach for pricing derivatives based on radial basis function (RBF) methods.

RBF methods are mesh-free numerical methods that exhibit a high, and for smooth problems, even exponential convergence order of the approximate solutions~\cite{Cheng,Kansa1,Madych,Rieger}. These methods are attractive from a computational point of view for higher-dimensional problems due to the simplicity of constructing interpolants using the radial basis. Moreover, RBF methods are able to achieve the desired accuracy using fewer computational nodes than, e.g., finite difference methods (FD)~\cite{Shcherbakov}, thus, reducing the storage requirements.
  
RBF methods were adapted for financial applications in the late 1990s and early 2000s by several authors~\cite{Fasshauer2,YCHon2,YCHon3,Pettersson}. However, they used a global RBF method with the basis functions globally supported over the entire computational domain. That approach resulted in a dense coefficient matrix of the system of linear equations that needs to be solved, therefore, preventing the method to be extended to higher dimensions because of the increase in computational complexity. To overcome this issue, several authors suggested to use localized RBF methods, such as the radial basis function partition of unity method (RBF--PUM)~\cite{Shcherbakov,Safdari,Shcherbakov2} and the radial basis function generated finite difference method (RBF--FD)~\cite{Slobodan, golbabai2017new, kadalbajoo2015efficient,kumar2015numerical, kadalbajoo2015application, kadalbajoo2013application}. 
These method modifications allow for a significant sparsification of the coefficient matrix, yielding a higher computational efficiency.  In fact, it was shown in~\cite{vonSydow} that \mbox{RBF--PUM} is more efficient than any other deterministic numerical method that relies on spatial discretization for higher-dimensional option pricing problems, e.g., for the Heston model, while RBF--FD at its current state of development is able to closely follow that result.
 
Thereby, we see a great potential of the localized RBF methods to be useful for pricing financial derivatives under models with several stochastic factors. In this paper, we demonstrate the performance of the RBF--PUM and RBF--FD methods on a set of pricing problems with multiple stochastic factors. The formulations of these problems are inspired by the BENCHOP project on stochastic and local volatility~\cite{BenchopMF}. For the problems with available semi-analytical solutions, we compute reference values and present convergence tests and computational performance comparisons, while for the other problems we provide solutions in the form of tables with values. Moreover, we discuss the parallelization potentials of our implementations and demonstrate the gains. Besides the positive highlights, we also draw attention to the limitations of the proposed methods and do not claim that these methods are suitable for solving problems with a very high dimensionality, e.g., larger than $10$. For those cases, most often, Monte Carlo methods are still the only way to obtain the solutions.

Nevertheless, problems of moderate dimensionality, such as a problem of pricing a basket option on five assets, e.g., an option on the DAX index after dimension reduction~\cite{DAX}, can be solved by both RBF--PUM and RBF--FD in a few seconds on an ordinary laptop~\cite{BenchopBasket}. Also, RBF--PUM has been successfully applied to a problem of valuing quanto CDS under a model with four stochastic factors~\cite{QuantoCDS}. 

The remainder of this paper is structured as follows. In Section~\ref{sec:Models}, we define four models with multiple stochastic factors under which we price a European call option. In Section~\ref{sec:Methods}, we develop the RBF--PUM and RBF--FD approaches that we use for estimating the option values numerically. In Section~\ref{sec:NumExper}, we present and discuss the numerical results. Finally, in Section~\ref{sec:Conclusions} we conclude the discussion on our findings.

%%%%%%%%%% --- MODELS --- %%%%%%%%%%%%
\section{Models with Multiple Stochastic Factors}\label{sec:Models}

Models with multiple stochastic factors allow for much better reproduction of market features than the standard Black--Scholes formulation, which is known for missing out on some important market features, such as fat tails of return distributions, as well as volatility smiles and skews. Therefore, various models with local volatilities, local stochastic volatilities, stochastic volatilities of volatilities, stochastic interest rates, and combinations of the above, have become increasingly popular since they are able to capture such market phenomena. 

In this section, we present four models with multiple stochastic factors that are often used for pricing options. Most of these model choices were inspired by~\cite{BenchopMF}.  In our setting we focus on European call options. However, the approach is not limited to this particular type of the payoff functions and can be easily extended to exotic payoffs as it is shown in~\cite{BenchopMF}.

We consider the models given a probability space $(\Omega,\mathcal{F},\mathds{Q})$ satisfying the standard assumptions, where $\Omega$ is the sample space, 
$(\mathcal{F}_t, t\geq 0)$ is the filtration under which the dynamics of the stochastic factors are adapted, and $\mathds{Q}$ is the risk neutral probability measure.

%Note that we pose the problems with respect to the time to maturity, i.e. we transform the time variable as $t=T-t$ that allows us to solve the problems forward in time.

% --- QLSV
\subsection{The Quadratic Local Stochastic Volatility Model}
The interest for local volatility models was triggered by the work of Dupire~\cite{Dupire}. Ever since, such models have become increasingly popular among practitioners, because they allow for capturing the volatility features that are observed in the market data, such as smiles and skews. The first reference to the quadratic local stochastic volatility (QLSV) relates to the paper by Andersen~\cite{Andersen2011}, where he applied the model for option pricing. 

The dynamics of the QLSV models read as follows
\begin{align}
\mathrm{d}S_t & =  rS_t\mathrm{d}t + \sqrt{V_t}f(S_t)\mathrm{d}W_t^s, \quad S_0=s, \label{qlsvSDE1} \\
\mathrm{d}V_t & =  \kappa(\eta-V_t)\mathrm{d}t + \sigma \sqrt{V_t}\mathrm{d}W_t^v, \quad V_0=v, \label{qlsvSDE2}
\end{align}
where $S_t$ is the stochastic asset price, $V_t$ is its stochastic volatility, $\sigma$ is the constant volatility of volatility, $\kappa$ is the speed of mean reversion of the volatility process, $\eta$ is the mean reversion level, $r$ is the risk-free interest rate,  $W_t^s$ and $W_t^v$ are correlated Wiener processes with constant correlation $\rho$, i.e. $\langle\mathrm{d}W_t^s, \mathrm{d}W_t^v\rangle = \rho \mathrm{d}t$, and $f(s)$ is a quadratic volatility function that takes the form
\begin{equation}
f(s) = \frac{1}{2}\alpha s^2 + \beta s + \gamma,
\end{equation}
with constant parameters $\alpha$, $\beta$, and $\gamma$. The standard Heston model~\cite{Heston} is a particular case of the QLSV model with \mbox{$\alpha =0$, $\beta=1$, and $\gamma=0$}.

By applying the It\^{o} lemma and the Feynman--Kac theorem, a pricing partial differential equation (PDE) for the QLSV model can be derived
\begin{equation}\label{qlsvPDE}
-\frac{\p u}{\p t} =  \frac{1}{2}vf(s)^2\frac{\p^2 u}{\p s^2} + \rho\sigma v f(s)\frac{\p^2 u}{\p s\p v} + \frac{1}{2}\sigma^2v\frac{\p^2 u}{\p v^2} + rs\frac{\p u}{\p s} + \kappa(\eta-v)\frac{\p u}{\p v} - ru, 
\end{equation}
subject to the terminal condition
\begin{equation}
u(T,s,v) = \max(s-K,0),
\end{equation}
where $K$ is the strike price and $s$ and $v$ are deterministic representations of the stochastic asset price and volatility processes, respectively. 

We set the following values for the model parameters
\begin{itemize}
\item Set 1:  $\alpha=0, \beta=1, \gamma=0$,
\item Set 2:  $\alpha=2, \beta=0, \gamma=0$,
\end{itemize}
while the other parameters are identical for both testing sets and selected as
$$
K=1, T=1, r=0, \kappa=2.58, \eta=0.043, \sigma=1,\rho=-0.36.
$$
The Feller condition~\cite{Feller51} is violated for both testing sets.
We choose three evaluation points at which we measure and report the option value results
$$
S_0 = 0.75, 1.00, 1.25; \qquad V_0 = 0.114.
$$
%

% --- SABR
\subsection{The SABR Model}
The stochastic alpha, beta, rho (SABR) model was developed by Hagan et al.~\cite{Hagan} to attempt to capture the volatility smile in derivative markets. The model is an extension of the constant elasticity of variance (CEV) model with an assumption of the volatility being stochastic. The stochastic asset and volatility processes are defined by the following dynamics
\begin{align}
\mathrm{d}S_t & =  V_tS_t^{\beta}\mathrm{d}W_t^s, \quad S_0 = s, \label{sabrSDE1} \\
\mathrm{d}V_t & =  \sigma V_t\mathrm{d}W_t^v, \quad V_0 = v, \label{sabrSDE2}
\end{align}
where $\beta$ is the elasticity parameter and the other parameters are defined as in \mbox{\eqref{qlsvSDE1}--\eqref{qlsvSDE2}}. 

By applying the It\^{o} lemma and the Feynman--Kac theorem we can derive a pricing PDE for the SABR model that takes the form
\begin{equation}\label{sabrPDE}
-\frac{\p u}{\p t} =  \frac{1}{2}v^{2} s^{2\beta}\frac{\p^2 u}{\p s^2} + \rho\sigma v^2 s^{\beta} \frac{\p^2 u}{\p s\p v} + \frac{1}{2}\sigma^2v^2\frac{\p^2 u}{\p v^2} - ru, 
\end{equation}
subject to the terminal condition
\begin{equation}
u(T,s,v) = \max(s-K,0).
\end{equation}

For valuing the option we assume the model parameters to have the following values
$$
K=1, T=1, r=0, \sigma=0.4, \beta=0.5.
$$
Here, we consider two test cases: zero-correlation between the asset and volatility for which there exists a semi-analytical solution~\cite{AntonovSABR} and nonzero-correlation between the asset and volatility for which some attempts to derive an approximate semi-analytical solution have been made~\cite{Islah} but the results are  still rather poor:
\begin{itemize}
\item Set 1:  $\rho=0$,
\item Set 2:  $\rho=-0.5$.
\end{itemize}
The three evaluation points at which we measure and report the computed option values are
$$
S = 0.75, 1.00, 1.25; \qquad V_0 = 0.200.
$$
The SABR model is typically used for valuing interest rate derivatives. 
%However, our parameter setting may not be representative for that purpose. Instead, we intentionally choose our parameters to have a uniform setup for all the models discussed in the paper, such that the performance of the methods can be more easily compared. 

% --- HHW
\subsection{The Heston--Hull--White Model}

The Heston--Hull--White (HHW) model is an extension of the Heston stochastic volatility model~\cite{Heston} that is augmented with a stochastic interest rate that follows the Hull--White process~\cite{HullWhite}. This is an important extension since the market interest rates are non-constant. Another notable property of the Hull--White model is that the interest rates can go negative, which nowadays happens in some economies, e.g. in 2015 the central bank of Sweden (Sveriges Riksbank) adopted a negative interest rate strategy for interbank lending to boost the economy of the country~\cite{Riksbank}.  

The three stochastic model factors are defined by the following dynamics
\begin{align}
\mathrm{d}S_t & =  R_tS_t\mathrm{d}t + \sqrt{V_t}S_t\mathrm{d}W_t^s, \quad S_0=s, \label{hhwSDE1} \\
\mathrm{d}V_t & =  \kappa(\eta-V_t)\mathrm{d}t + \sigma_v \sqrt{V_t}\mathrm{d}W_t^v, \quad V_0=v, \label{hhwSDE2} \\
\mathrm{d}R_t & = a(b-R_t)\mathrm{d}t + \sigma_r\mathrm{d}W_t^r, \quad R_0 = r, \label{hhwSDE3}
\end{align}
where $R_t$ is the stochastic interest rate, $a$ is the speed of mean reversion of the interest rate process, $b$ is its mean reversion level, $\sigma_r$ is its volatility,  $W_t^s$, $W_t^v$,  and $W_t^r$ are correlated Wiener processes with constant correlation $\rho_{ij}$, i.e. $\langle \mathrm{d}W_t^i, \mathrm{d}W_t^j\rangle = \rho_{ij} \mathrm{d}t$, $i, j \in \{s,v,r\}$, and the other parameters are defined as in \eqref{qlsvSDE1}--\eqref{qlsvSDE2}. 

We can apply the It\^{o} lemma and the Feynman--Kac theorem to derive a pricing PDE for the HHW model that takes the following form
\begin{align}\label{hhwPDE}
-\frac{\p u}{\p t} = &  \frac{1}{2}vs^2\frac{\p^2 u}{\p s^2} + \frac{1}{2}\sigma_v^2v\frac{\p^2 u}{\p v^2}  + \frac{1}{2}\sigma_r^2\frac{\p^2 u}{\p r^2} + \nonumber \\
                           & \rho_{sv}\sigma_v vs\frac{\p^2 u}{\p s\p v} + \rho_{sr}\sigma_r \sqrt{v} s\frac{\p^2 u}{\p s\p r} + \rho_{vr}\sigma_v\sigma_r \sqrt{v}\frac{\p^2 u}{\p v\p r} + \nonumber \\
                           & rs\frac{\p u}{\p s} + \kappa(\eta-v)\frac{\p u}{\p v} + a(b-r)\frac{\p u}{\p r} - ru, 
\end{align}
subject to the terminal condition
\begin{equation}
u(T,s,v,r) = \max(s-K,0).
\end{equation}

The property of the interest rate to take negative values introduces a numerical issue when solving the pricing PDE. Some methods such as FD methods encountered and reported spurious oscillations in the numerical solution~\cite{intHoutHHW}. To alleviate the spurious oscillations an upwind FD scheme can be constructed and used when $r<0$. 

The model parameter values that we use for the numerical experiments are
\begin{align*}
& K=1, T=1, \kappa=0.50, \eta=0.04, \sigma_v=0.25, \sigma_r=0.09, \rho_{sv} = -0.9, \\
& \rho_{sr} = 0.6, \rho_{vr} = -0.7, a=0.08, b=0.1,
\end{align*}
and the three evaluation points are
$$
S = 0.75, 1.00, 1.25; \qquad V_0 = 0.040; \qquad R_0 = 0.100.
$$ 
%

% --- Heston--CIR
\subsection{The Heston--Cox--Ingersoll--Ross Model}

The Cox--Ingersoll--Ross (CIR) interest rate model~\cite{CIR} is another model for describing the stochastic nature of interest rates. We combine the CIR stochastic interest rate model with the Heston stochastic volatility model to obtain a three-factor structure that we call the Heston--Cox--Ingersoll--Ross (HCIR) model. A notable property and the difference of the CIR model compared to the Hull--White model is that the interest rates cannot be negative.

The dynamics of the stochastic processes read as follows
\begin{align}
\mathrm{d}S_t & =  R_tS_t\mathrm{d}t + \sqrt{V_t}S_t\mathrm{d}W_t^s, \quad S_0=s, \label{hcirSDE1}\\
\mathrm{d}V_t & =  \kappa(\eta-V_t)\mathrm{d}t + \sigma_v \sqrt{V_t}\mathrm{d}W_t^v, \quad V_0=v, \label{hcirSDE2}\\
\mathrm{d}R_t & = a(b-R_t)\mathrm{d}t + \sigma_r\sqrt{R_t}\mathrm{d}W_t^r, \quad R_0=r. \label{hcirSDE3}
\end{align}

In order to derive a pricing PDE we commit to the standard procedure of applying the It\^{o} lemma and the Feynman--Kac theorem. The obtained PDE takes the form
\begin{align}\label{hcirPDE}
-\frac{\p u}{\p t} = &  \frac{1}{2}vs^2\frac{\p^2 u}{\p s^2} + \frac{1}{2}\sigma_v^2v\frac{\p^2 u}{\p v^2}  + \frac{1}{2}\sigma_r^2r\frac{\p^2 u}{\p r^2} + \nonumber \\
                           & \rho_{sv}\sigma_v vs\frac{\p^2 u}{\p s\p v} + \rho_{sr}\sigma_r \sqrt{v}\sqrt{r}s\frac{\p^2 u}{\p s\p r} +
                               \rho_{vr}\sigma_v\sigma_r \sqrt{v}\sqrt{r}\frac{\p^2 u}{\p v\p r} + \nonumber \\
                           & rs\frac{\p u}{\p s} + \kappa(\eta-v)\frac{\p u}{\p v} + a(b-r)\frac{\p u}{\p r} - ru, 
\end{align}
subject to the terminal condition
\begin{equation}
u(T,s,v,r) = \max(s-K,0).
\end{equation}

Here we use the same parameter values as for the HHW model, that is,
\begin{align*}
& K=1, T=1, \kappa=0.50, \eta=0.04, \sigma_v=0.25, \sigma_r=0.09, \rho_{sv} = -0.9, \\
& \rho_{sr} = 0.6, \rho_{vr} = -0.7, a=0.08, b=0.1,
\end{align*}
and the three evaluation points are 
$$
S = 0.75, 1.00, 1.25; \qquad V_0 = 0.040; \qquad R_0 = 0.100.
$$
%

%%%%%%%%%% --- METHODS --- %%%%%%%%%%%%
\section{Localized Radial Basis Function Methods}\label{sec:Methods}
RBF methods were first used to approximate solutions of partial differential equations by Kansa \cite{Kansa1,Kansa2} in the early 1990s. Since then, the methods have gained in popularity and have been applied to various types of problems of mathematical physics~\cite{Ferreira,FlyerLehto}, glaciology~\cite{Ahlkrona,CG}, quantum physics~\cite{Kormann2013}, and computational finance~\cite{Fasshauer2,YCHon3,Pettersson}. RBF methods exhibit several very attractive properties, such as a high order convergence of the approximated solution~\cite{Rieger,Rieger2,RBFLS} and flexibility with respect to the computational domain geometry~\cite{Ahlkrona,RBFLS}. However, the main issue with the standard global formulation~\cite{Fasshauer} is a dense coefficient matrix. In this paper, we address this issue by utilizing the mentioned localized approaches. It has been shown in the BENCHOP project \cite{vonSydow}, that localized RBF approximations make efficient numerical methods for partial differential equations and that they have a great potential for solving multi-dimensional problems. Here, we develop two localized RBF techniques, namely RBF--PUM and \mbox{RBF--FD}. 

Since the problems \eqref{qlsvPDE}, \eqref{sabrPDE}, \eqref{hhwPDE}, and \eqref{hcirPDE} are stated in infinite domains, in order to perform numerical simulations, we need to truncate the domains and select appropriate boundary conditions at the boundaries imposed by the truncations. We denote the truncated domain as $\hat\Omega$. For the introduced two-factor models we define $\hat\Omega:=[s_{\text{min}}, s_{\text{max}}]\times[v_{\text{min}}, v_{\text{max}}]$, while for the three-factor models \mbox{$\hat\Omega:=[s_{\text{min}}, s_{\text{max}}]\times[v_{\text{min}}, v_{\text{max}}]\times[r_{\text{min}}, r_{\text{max}}]$}.

\begin{figure}[H]
    \centering
    \begin{subfigure}[b]{0.45\textwidth}
        \includegraphics[width=\textwidth]{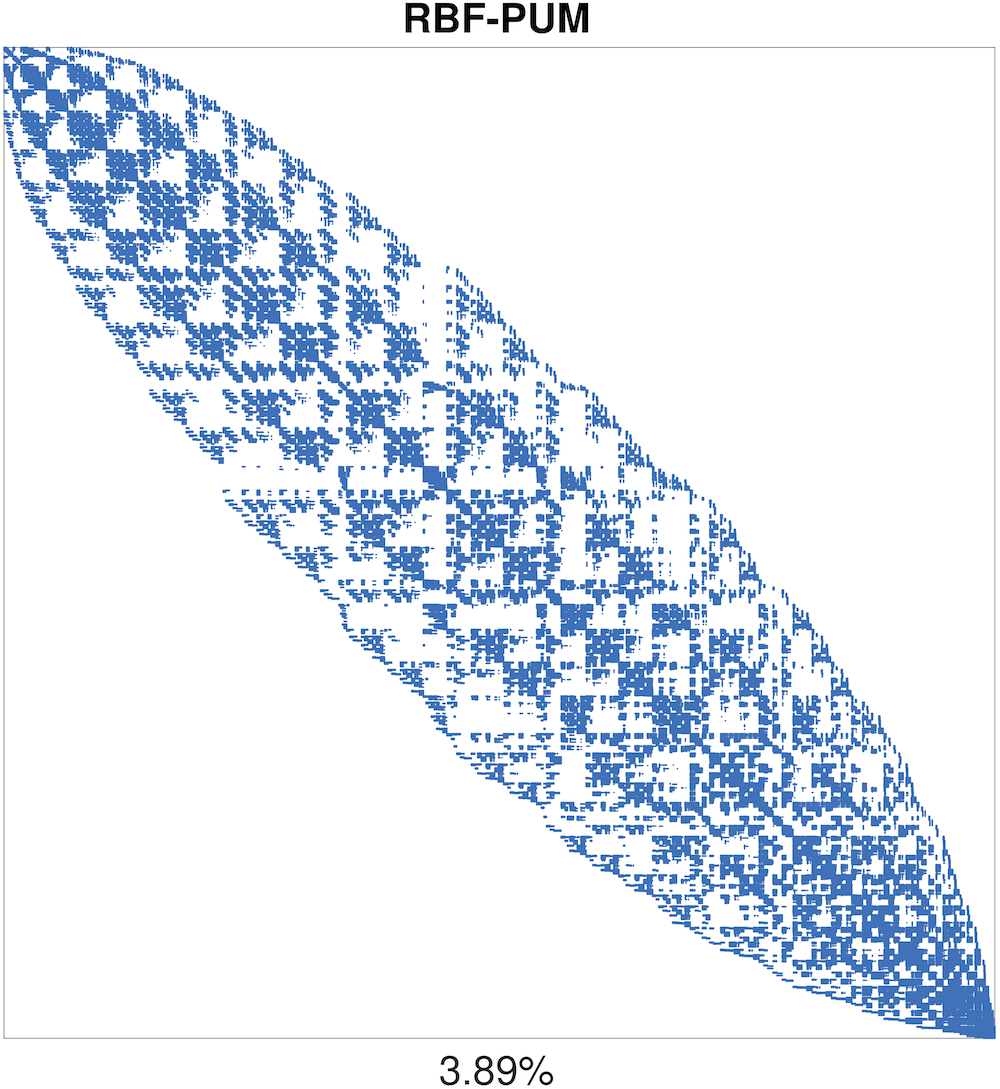}
    \end{subfigure}
    \hspace{0.6cm}
    \begin{subfigure}[b]{0.45\textwidth}
        \includegraphics[width=\textwidth]{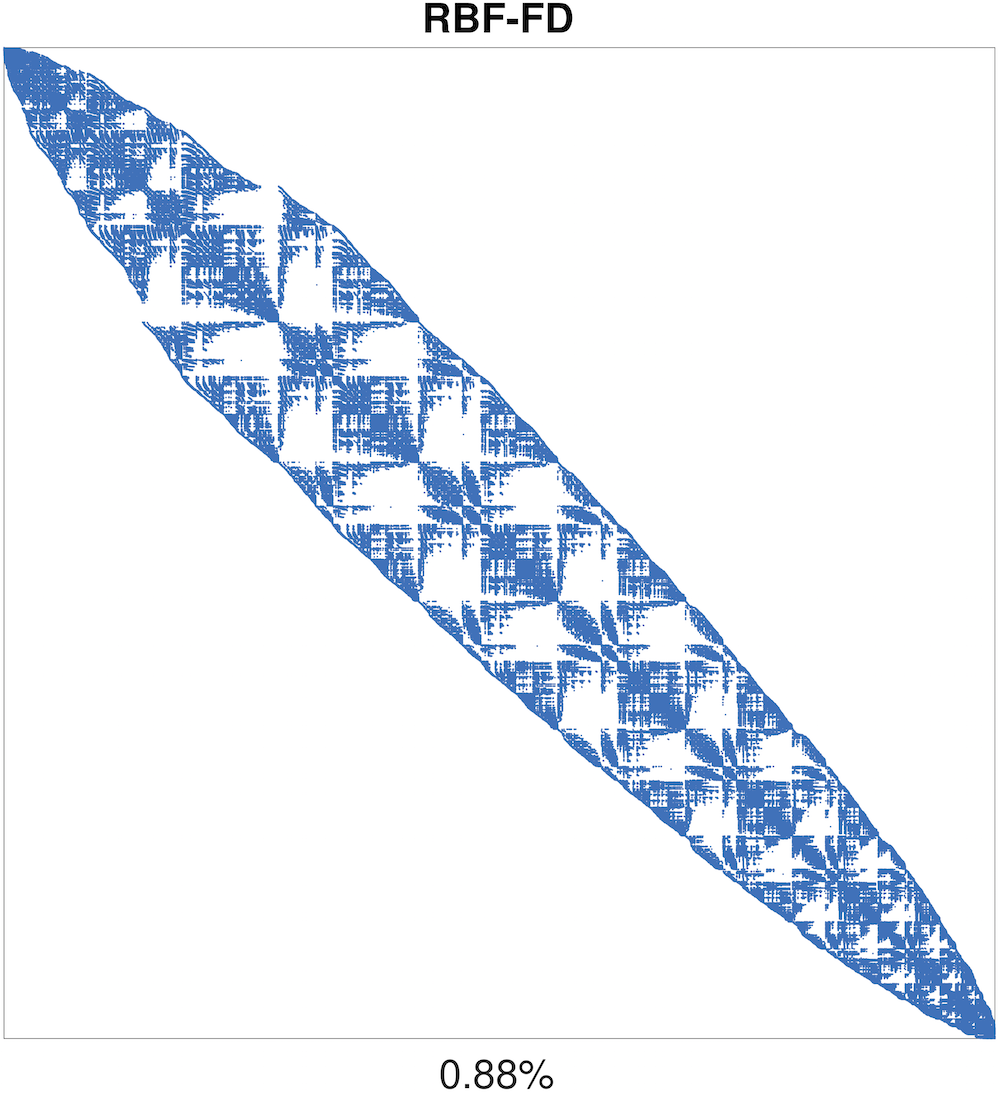}
    \end{subfigure}
 \caption{An example of the sparsity structures of the differentiation matrices obtained from the discretization of the HHW problem for RBF--PUM (\emph{left}) and RBF--FD (\emph{right}).}\label{SparseStr}
\end{figure}

%\begin{figure}[H]
%\begin{center}
%\subfigure[RBF--PUM]{
%\resizebox*{6cm}{!}{\includegraphics{spyRBFPUM.png}}}\hspace{25pt}
%\subfigure[RBF--FD]{
%\resizebox*{6cm}{!}{\includegraphics{spyRBFFD.png}}}
%\caption{An example of the sparsity structures of the differentiation matrices obtained from the discretization of the HHW problem.}
%\label{SparseStr}
%\end{center}
%\end{figure}

For convenience, we rewrite equations \eqref{qlsvPDE}, \eqref{sabrPDE}, \eqref{hhwPDE}, and \eqref{hcirPDE} in a shorthand form and augment them with the boundary conditions
\begin{align}
\frac{\p u}{\p t} + \mathcal{L}u(t,\ul{x}) &= 0, \quad \ul{x} \in \hat \Omega, \label{genPDE} \\
\mathcal{B}u(t,\ul{x}) &= f(t,\ul{x}), \quad \ul{x} \in \p \hat \Omega, \label{genBC}
\end{align}
where $\mathcal{L}$ is the corresponding differential operator and $\mathcal{B}$ is the boundary differential operator, while $f(t,\ul{x})$ is the forcing function and $\ul{x}=[s,v]$ or $\ul{x}=[s,v,r]$ depending on the dimensionality of the model.

In Figure \ref{SparseStr}, we demonstrate the sparse structure of the discretized differential operator for the HHW model, that follows from the approximations by our methods after the reverse Cuthill--McKee reordering~\cite{Cuthill1969,George1981} to reduce the bandwidth.

\begin{figure}[H]
\captionsetup[subfigure]{justification=centering}
    \centering
    \begin{subfigure}[b]{0.32\textwidth}
        \includegraphics[width=\textwidth]{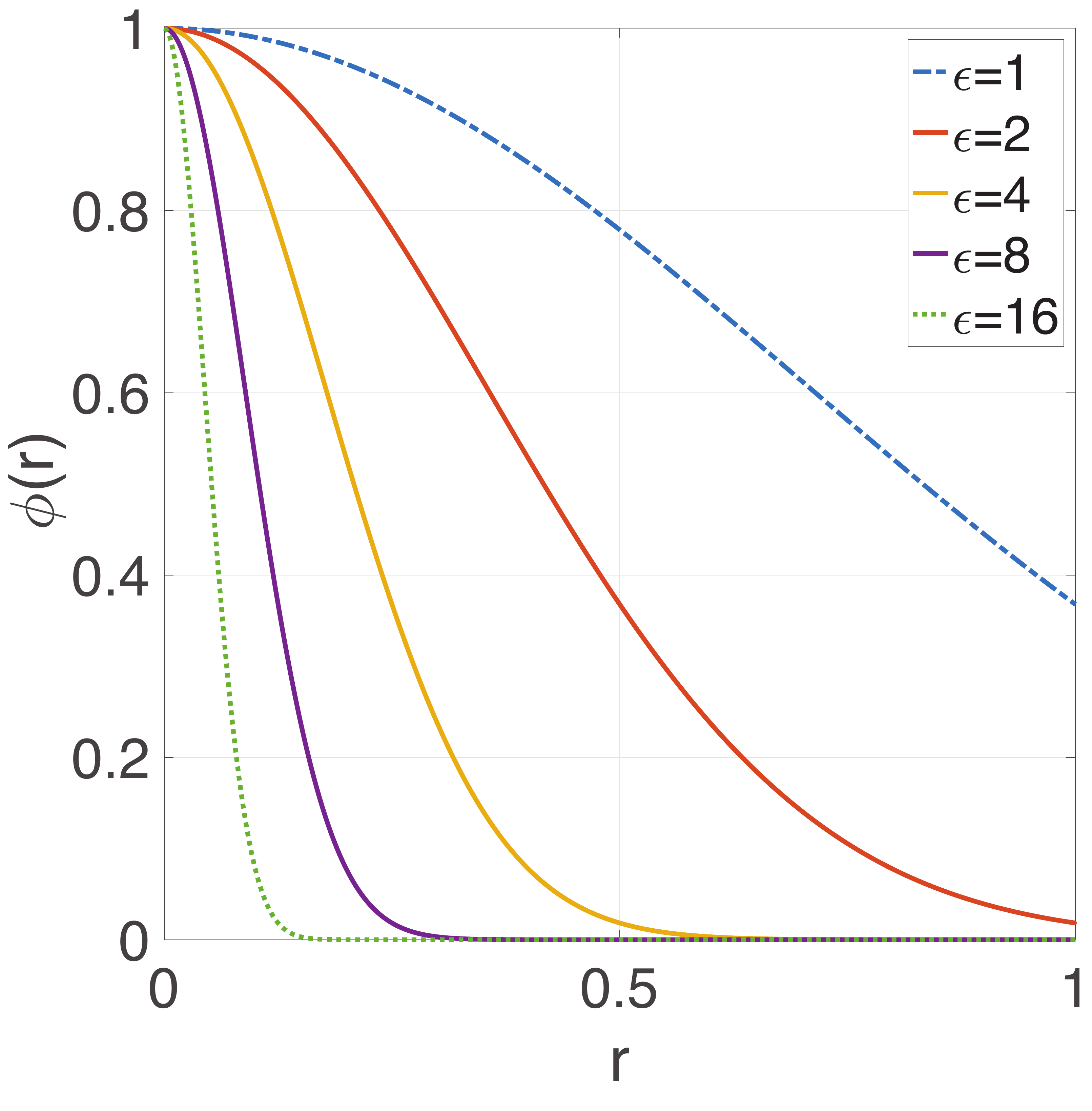}
        \caption{Gaussian}
    \end{subfigure}
    \begin{subfigure}[b]{0.32\textwidth}
        \includegraphics[width=\textwidth]{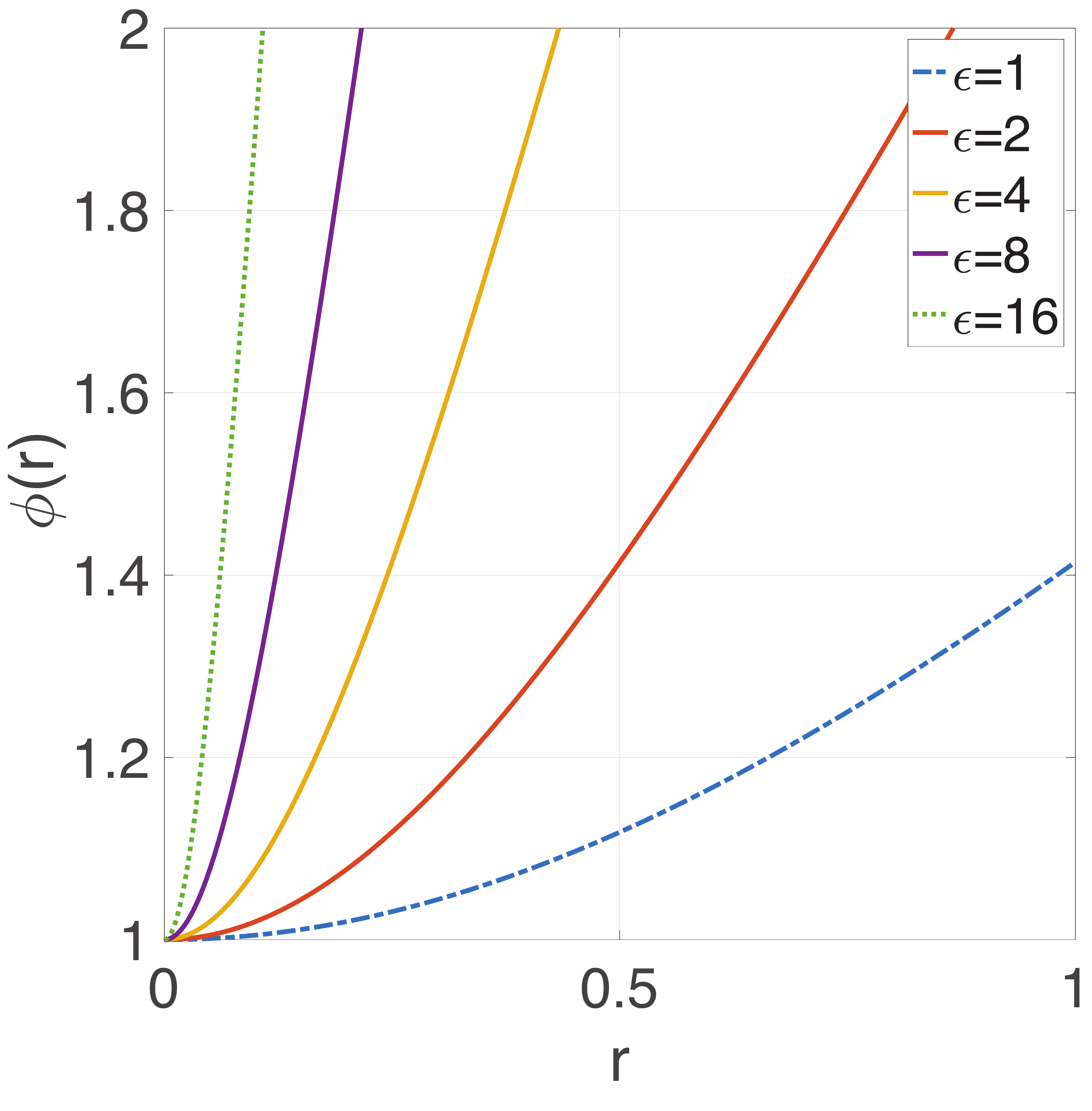}
        \caption{Multiquadric}
    \end{subfigure}
     \begin{subfigure}[b]{0.32\textwidth}
        \includegraphics[width=\textwidth]{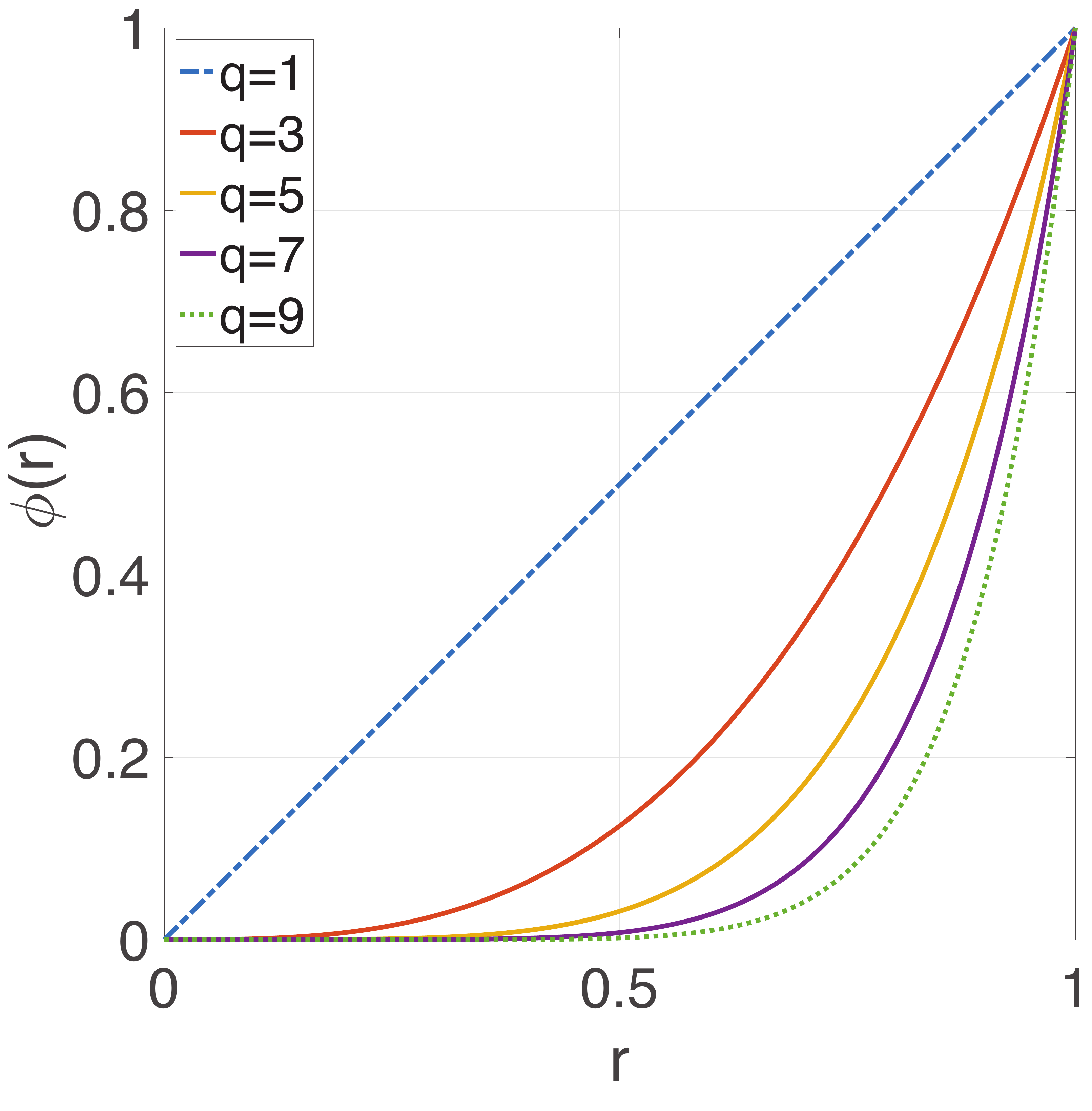}
        \caption{Polyharmonic spline}
    \end{subfigure}
 \caption{Commonly used basis functions with respect to the value of the shape parameter $\varepsilon$ or the polyharmonic spline degree $q$.}\label{BasisFunc}
\end{figure}

In Table \ref{TabRBF} and Figure \ref{BasisFunc} we present a set of typical choices of RBFs used to develop numerical methods and illustrate how the shape parameter affects the scaling of the functions.
\begin{table}[!ht]
\begin{center}
\caption{Commonly used radial basis functions, where $\varepsilon\in \mathds{R}^+$ is the shape parameter and $q\in\{2m-1, m \in \mathds{N}\}$ is the polyharmonic spline degree.}
\begin{tabular}{ l  c  c  c  r  }
%\hline\hline
RBF & & &  & $\phi(r)$   \\ \hlinewd{1pt}
 Gaussian (GA) &  & &  &  $\exp{(-\varepsilon^2r^2)}$ \\
 Multiquadric (MQ) &  & &  & $\sqrt{1+\varepsilon^2r^2}$ \\
 Inverse Multiquadric (IMQ) & & &  & $1/\sqrt{1+\varepsilon^2r^2}$ \\
 Inverse Quadratic (IQ) & & &  & $1/(1+\varepsilon^2r^2)$ \\
 Polyharmonic Spline (PHS) & & &  & $r^q$\\
\hlinewd{1pt}
%\hline\hline
\end{tabular}
\label{TabRBF}
\end{center}
\end{table}
%
%
%\begin{figure}[H]
%\begin{center}
%\subfigure[Gaussian]{\resizebox*{4.5cm}{!}{\includegraphics{gs.eps}}}
%\subfigure[Multiquadric]{\resizebox*{4.5cm}{!}{\includegraphics{mq.eps}}}
%\subfigure[Polyharmonic spline]{\resizebox*{4.5cm}{!}{\includegraphics{phs.eps}}}
%\caption{Commonly used basis functions with respect to the value of the shape parameter $\varepsilon$ or the polyharmonic spline degree $q$.}
%\label{BasisFunc}
%\end{center}
%\end{figure}

%

In the following parts of this section, we specify the details of the RBF methods and discuss their properties. Additionally, we present the main implementation steps in the form of pseudocode algorithms in the appendices of this paper. 

% --- RBF--PUM
\subsection{RBF--PUM}

In order to construct an RBF--PUM approximation we start by defining an open cover $\{\Omega_j\}_{j=1}^{P}$ of the computational domain $\hat\Omega\subset \mathds{R}^{d}$ such that
\begin{equation}
  \hat \Omega \subseteq \bigcup_{j=1}^{P} \Omega_j.
\end{equation}
We select the patches $\Omega_j$ to be of a spherical form. Inside each patch a local RBF approximation of the solution $u$ is defined as
\begin{equation}\label{localRBF}
  \tilde u_j(t,\ul{x})= \sum_{i=1}^{n_j}\lambda_i^j(t) \phi(\| \ul{x} - \ul{x}_i^j \|),
\end{equation}
\noindent where $n_j$ is the number of computational nodes belonging to the patch $\Omega_j$, $\phi( \|\ul{x} - \ul{x}_i^j \|)$ is the $i$-th basis function centered at $\ul{x}_i^j$, which is the $i$-th local node in the $j$-th patch $\Omega_j$, and $\lambda_i^{j}(t)$ are the unknown coefficients.

In addition to the patches, we construct partition of unity weight functions
$w_j(\ul{x}),\, j = 1, \ldots, P$, subordinated to the open cover, such that
\begin{equation}
  \sum_{j=1}^{P} w_j(\ul{x}) = 1, \quad \forall \ul{x}\in \hat\Omega.
\end{equation}
Functions $w_j(\ul{x})$ can be obtained, e.g., by Shepard's method, \cite{Shepard}, from compactly supported generating functions $\varphi_j(\ul{x})$
\begin{equation}\label{PUweight}
  w_j(\ul{x}) = \frac{\varphi_j(\ul{x})}{\sum_{i=1}^{P} \varphi_i(\ul{x})}, \quad j=1,\ldots,P, \quad \forall \ul{x}\in \hat\Omega.
\end{equation}
The generating functions $\varphi_j(\ul{x})$ must fulfil some smoothness requirements. For instance, for the problems considered in this paper
they should be at least $C^{2}(\mathds{R}^{d})$. To proceed, as a suitable candidate for $\varphi_j(\ul{x})$ we choose compactly supported Wendland functions~\cite{Wendland},
\begin{equation}
  \varphi(r) = (4r+1)(1-r)^4_{+}, \quad r \in \mathds{R},
\end{equation}
\noindent with $\text{supp}(\varphi(r)) = \mathds{B}^{d}(0, 1)$, where $\mathds{B}^d(0,1)$ is a unit $d$-dimensional ball centered at the origin. In order to map the generating function to the patch $\Omega_j$ with centre $\ul{c}_j$ and radius $\rho_j$, we shift and scale it as
\begin{equation}
 \varphi_{j}(\ul{x}) = \varphi_j \left( \frac{\|\ul{x}- \ul{c}_j\|}{\rho_j}\right), \quad \forall \ul{x} \in \hat\Omega.
\end{equation}
Further we combine the local RBF approximations with the partitions of unity weight functions and obtain a global RBF--PUM solution $\tilde u(\ul{x})$ as
\begin{equation}\label{RBFPUMapprox}
 \tilde u(t,\ul{x}) = \sum_{j=1}^{P}w_j(\ul{x}) \tilde u_j(t,\ul{x}).
\end{equation}

In order to attack the problem \eqref{genPDE}--\eqref{genBC} numerically by RBF--PUM we need to scatter $N$ nodes $\bold{x} = \{\ul{x}_1,\ldots,\ul{x}_N\}$ in $\hat\Omega$. Without loss of generality, we assume that the first $N_I$ nodes belong to the interior of $\hat\Omega$ and the remaining $N_B = N-N_I$ nodes belong to its boundary. We seek the solution to \eqref{genPDE}--\eqref{genBC} in the form of \eqref{RBFPUMapprox}. Collocating \eqref{RBFPUMapprox} at the nodes we obtain the following linear system of equations
\begin{equation}\label{Lambda_setup}
  \begin{bmatrix}
    A\\
    0
  \end{bmatrix} \frac{\mathrm{d}\vec{\lambda}}{\mathrm{d}t} 
  +
  \begin{bmatrix}
    L\\
    B
  \end{bmatrix} \vec{\lambda}
  :=
  \begin{bmatrix}
    A_{II} & A_{BI}\\
    0 & 0
  \end{bmatrix} \frac{\mathrm{d}}{\mathrm{d}t}
  \begin{bmatrix}
    \vec{\lambda}_{I} \\
    \vec{\lambda}_{B}
  \end{bmatrix}
  +
  \begin{bmatrix}
    L_{II} & L_{BI}\\
    B_{IB} & B_{BB}
  \end{bmatrix}
  \begin{bmatrix}
    \vec{\lambda}_{I} \\
    \vec{\lambda}_{B}
  \end{bmatrix}
  =
  \begin{bmatrix}
    0 \\ 
    \vec{f}_B
  \end{bmatrix},
\end{equation}
where $L$ and $B$ are the discrete representation of the differential operator and boundary differential  operator, respectively, $A$ is the interpolation matrix, $\vec{\lambda}_{I} = [\lambda_1,\ldots,\lambda_{N_I}]^{T}$, $\vec{\lambda}_{B} = [\lambda_{N_I+1},\ldots,\lambda_{N}]^{T}$, and $\vec{f}_{B}(t) = [f(t,\ul{x}_{N_{I} +1}),\ldots,f(t,\ul{x}_{N})]^{T}$.

However, using the interpolation relation between $\vec{u}$ and $\vec{\lambda}$ on \eqref{RBFPUMapprox} at the nodes we get the following system of equations 
\begin{equation}
  \vec{u} = \mathcal{A} \vec{\lambda}, \qquad 
  \mathcal{A} := 
  \begin{bmatrix}
    A_{II} & A_{BI}\\
    A_{IB} & A_{BB}
  \end{bmatrix},
\end{equation}
which can be used to express the coefficients $\vec{\lambda}$ in terms of the function values $\vec{u}$ as
\begin{equation}\label{U_Lambda}
  \vec{\lambda} = \mathcal{A}^{-1} \vec{u}.
\end{equation}
For the smooth basis functions from Table~\ref{TabRBF} the matrix $\mathcal{A}$ is non-singular, i.e., $\mathcal{A}^{-1}$ exists~\cite{Micchelli}. Then using the relationship \eqref{U_Lambda} we can transform the problem \eqref{Lambda_setup} of finding the coefficients $\vec{\lambda}$ to a problem of finding the function values directly.
It has been shown~\cite{Driscoll,Larsson7,Schaback2} that for smooth RBFs the magnitude of the coefficients $\vec{\lambda}$ becomes unbounded as $\varepsilon\to 0$, while the values $\vec{u}$ remain well-behaved. Therefore, we prefer to express the problem in terms of the nodal function values $\tilde{u}$. 
\begin{equation}\label{U_setup}
  \begin{bmatrix}
    A_{II} & A_{BI}\\
    0 & 0
  \end{bmatrix} \mathcal{A}^{-1}\frac{\mathrm{d}}{\mathrm{d}t}
  \begin{bmatrix}
    \vec{u}_{I} \\
    \vec{u}_{B}
  \end{bmatrix}
  +
  \begin{bmatrix}
    L_{II} & L_{BI}\\
    B_{IB} & B_{BB}
  \end{bmatrix} \mathcal{A}^{-1}
  \begin{bmatrix}
    \vec{u}_{I} \\
    \vec{u}_{B}
  \end{bmatrix}
  =
  \begin{bmatrix}
    0 \\ 
    \vec{f}_B
  \end{bmatrix},
\end{equation}
where $\vec{u}_{I}(t) = [\tilde u(t,\ul{x}_1),\ldots,\tilde u(t,\ul{x}_{N_{I}})]^{T}$ and $\vec{u}_{B}(t) = [\tilde u(t,\ul{x}_{N_{I} +1}),\ldots, \tilde u(t,\ul{x}_{N})]^{T}$.

The formulation \eqref{U_setup} is a system of ordinary differential equations, which can, in principle, be numerically solved by any suitable method. We refer to Section~\ref{sec:BDF2} for the time integration details. Moreover, the algorithm for the RBF--PUM solver is presented in Appendix \ref{A:rbfpum}.

The RBF--PUM approximation in the given form allows to maintain accuracy similar to that of the global method while significantly reducing the computational effort~\cite{Shcherbakov,Ahlkrona} thanks to the sparse structure of the coefficient matrix that additionally enables the use of sparse operations in environments that support such functionality, e.g., in \textsc{MATLAB}. Moreover, RBF--PUM is well suited for parallel implementation since the computation of the local matrices is parallelizable.  

%The sparse structure of the RBF--PUM coefficient matrix for the Heston--Hull--White problem is displayed in Figure~\ref{SparseStr} in the left panel as an example

% --- RBF--FD
\subsection{RBF--FD}
In order to construct an RBF--FD approximation we start by scattering $N$ nodes accross the computational domain $\hat\Omega\subset \mathds{R}^{d}$. For each node $\ul{x}_j$, we define subsets $\{\Omega_j\}_{j=1}^N$ consisting of $n_j-1$ neighboring nodes and $\ul{x}_j$ itself, and consider them as stencils of size $n_j$. 
%In Figure \ref{fig:222}, we show an example of an adapted 2D non-uniform node layout with stencils of different sizes, that would be suitable for pricing e.g. a European basket option consisting of two assets, under the Black-Scholes model. 

The differential operator $\mathcal{L}$ defined in (\ref{genPDE})  is approximated in every node  $\ul{x}_j$ as
\begin{equation}
\mathcal{L}u(\ul{x}_j)\approx\sum_{i=1}^{n_j}{w}_{i}^{j}u_i^{j}\equiv Wu(\ul{x}_j),\quad j=1,\ldots,N,
\label{eq:RBF--FD}
\end{equation}
where $u_i^{j}=u(\ul{x}_i^j)$, and $\ul{x}_i^j$ is a locally indexed node as introduced before.

When it comes to the standard RBF--FD method introduced in~\cite{Tolstykh2, FoWri06}, the weights ${w}_i^j$ are calculated by enforcing (\ref{eq:RBF--FD}) to be exact for RBFs centered at each of the nodes in $\Omega_j$ yielding

\begin{equation}\label{eq:D}
{\small{
\left[\begin{array}{cccc}
\phi(\|\ul{x}_1^{j}-\ul{x}_1^{j}\|) & \ldots & \phi(\|\ul{x}_1^{j}-\ul{x}_{n_j}^{j}\|)\\
\vdots & \ddots & \vdots\\
\phi(\|\ul{x}_{n_j}^{j}-\ul{x}_1^{j}\|) & \ldots & \phi(\|\ul{x}_{n_j}^{j}-\ul{x}_{n_j}^{j}\|)
\end{array}\right]
\left[\begin{array}{c}
{w}_1^{j}\\
\vdots\\
{w}_{n_j}^{j}
\end{array}\right]=
\left[\begin{array}{c}
\mathcal{L}\phi(\|\ul{x}_{j}-\ul{x}_1^{j}\|)\\
\vdots \\
\mathcal{L}\phi(\|\ul{x}_{j}-\ul{x}_{{n_j}}^{j}\|)
\end{array}\right].}}
\end{equation}

From the theory on RBF interpolation, (\ref{eq:D}) forms a nonsingular system which means that a unique set of weights can be computed and assembled into a differentiation matrix $L$. Since $n_j \ll N$ the resulting differentiation matrix is sparse, making the memory cost ${\cal{O}}(N)$ compared to ${\cal{O}}(N^2)$ for global approximations with RBFs. Moreover, besides the freedom of node placement, the method is also featured with freedom of choice of the stencil size $n_j$ for each node $\ul{x}_j$ in the computational domain $\hat\Omega$, which can be used to control approximation accuracy in different parts of the domain.

%\begin{figure}[H]
%\centering
% \includegraphics[width=0.5\textwidth]{Figures/rbffd_gridsten.eps}
%\caption{Examples of nearest neighbor based stencils used for approximating the spatial differential operator $\mathcal{L}$ on a non-uniform node layout adapted for solving the two-dimensional Black-Scholes equation. The central node of each displayed stencil is denoted by a black cross mark, while the nodes that belong to each of the displayed stencils are shown in green. The nodes that require boundary conditions treatment are denoted by a yellow triangle and red squares.}
%\label{fig:222}
%\end{figure}

Many different types of RBFs have been considered for approximating similar differential operators in recent history and some of them are listed in Table \ref{TabRBF} and shown in Figure \ref{BasisFunc}. The main issue following this kind of approximations is that the linear systems of equations that need to be solved in order to obtain the weights $w_i^j$ are most often ill-conditioned. The shape parameter $\varepsilon$, featured in most of the RBFs needs to be picked very carefully and sometimes even that may not be enough to ensure a stable approximation. 

In several works in the past decade~\cite{Slobodan,FlyerLehto,davydov2011adaptive, fornberg2011stabilization, Larsson2, RBFGA,flyer2016enhancing}, it was demonstrated that the stencil approximation can be stabilized by adding low-order polynomials together with RBFs into the presented interpolation. Still, the problem of choosing the optimal shape parameter, which is thoroughly examined for option pricing problems in~\cite{Slobodan}, remains unsolved for the general case. Nevertheless, recent developments~\cite{FoFly1, Bayona}, showed that the RBF--FD approximation can be greatly improved by using high-order polynomials together with RBFs. It appears that, in this setting, the polynomial degree, instead of the RBF, dictates the rate of convergence. This finding suggests choosing piecewise smooth polyharmonic splines (PHS) as RBFs since they are not featured with a shape parameter. Still, it seems that RBFs do contribute to reduction of approximation errors and, therefore, are necessary for having both stable and accurate approximation.

Therefore, the linear system that we solve to get the differentiation weights for each node in our problems is of the following form
\begin{equation}\label{eq:D2}
{\small{
\left[\begin{array}{cc}
A & P^T \\
P & 0 \\
\end{array}\right]
\left[\begin{array}{c}
{\ul{w}}_j\\
{\ul{\gamma}}_j\\
\end{array}\right]=
\left[\begin{array}{c}
\mathcal{L}\phi(\|\ul{x}^{j}-\ul{x}_1^{j}\|)\\
\vdots \\
\mathcal{L}\phi(\|\ul{x}^{j}-\ul{x}_{{n_j}}^{j}\|)\\
\mathcal{L}p_1(\ul{x}_j)\\
\vdots\\
\mathcal{L}p_{m_j}(\ul{x}_j)
\end{array}\right],}}
\end{equation}
where $A$ is the RBF matrix and $\ul{w}_j$ is the vector of differentiation weights, both shown on the left-hand side of equation (\ref{eq:D}). $P$ is the matrix of size $m_j \times n_j$ that contains all monomials up to order $q_p$ (corresponding to $m_j$ monomial terms) that are evaluated in each node $\ul{x}_i^j$ of the stencil $\Omega_j$ and $0$ is a zero square matrix of size $m_j \times m_j$. Furthermore, $\ul{\gamma}_j$ is a vector of dummy weights that should be discarded and $\{p_1, p_2, \ldots, p_{m_j}\}$ is an array of monomial functions indexed by their position relative to the total number of monomial terms $m_j$, such that it contains all the combinations of monomial terms up to degree $q_p$.

It is worth noting that compared to standard FD discretizations, where differential operators are approximated only on one-dimensional Cartesian grids, which means that high-dimensional operators need to be discretized separately in each direction, in the RBF--FD approximations, dimensionality does not play a role in the difficulty of this problem. Moreover, when it comes to the boundary nodes and the ones that are close to it, the nearest neighbor based stencils automatically deform, therefore requiring no special treatment for computing the differentiation weights in those areas. The only information that is required for the approximation of differential operators are the Euclidian distances between the nodes. This means that (\ref{eq:D}) represents a way to approximate a differential operator in any number of dimensions. Furthermore, it is true that the FD weights can be directly derived and that for the RBF--FD case one has to solve a small linear system for each node to obtain them, but we need to stress that this task (corresponding to the \texttt{for} loop at lines $4$--$11$ of Algorithm \ref{AlgRBFFD} in Appendix \ref{A:rbffd}) is perfectly parallelizable and that this extra cost can be justified by the great features of the method presented so far. Finally, if a special boundary condition needs to be imposed, that can be done by simply replacing the operator $\mathcal{L}$ with an appropriate boundary operator $\mathcal{B}$ when approximating the weights for the nodes that belong to the boundary.

After the weights are computed and stored in the differentiation matrix, approximation of  (\ref{genPDE}) and (\ref{genBC}) looks as the following semi-discrete equations
\begin{equation}
\label{dRBFFD}
\left[\begin{array}{cc}
E_{II} & 0_{IB} \\
0_{BI} & 0_{BB} \\
\end{array}\right]
\frac{\mathrm{d}}{\mathrm{d} t}
\left[\begin{array}{c}
\vec{u}_I\\
\vec{u}_B\\
\end{array}\right] =
\left[\begin{array}{cc}
L_{II} & L_{IB} \\
B_{BI} & B_{BB} \\
\end{array}\right]
\left[\begin{array}{c}
\vec{u}_I\\
\vec{u}_B\\
\end{array}\right],
\end{equation}
which can be integrated in time as described in the following subsection. Moreover, the algorithm for the RBF--FD solver is presented in Appendix \ref{A:rbfpum}.

% --- Time-stepping scheme
\subsection{Integration in Time}\label{sec:BDF2}
\label{ssec:BDF2}
For the time discretization we select the unconditionally stable second order backward differentiation formula (BDF-2) \cite[p.~401]{Hairer}. 
The BDF-2 scheme involves three time levels. To initiate the method, often the BDF-1 scheme (Euler backward) is used for the first time step. Thus, two different matrices need to be factorized. In order to avoid this we use the BDF-2 scheme as described in~\cite{Larsson3}.

We split the time interval $[0,T]$ into $N_{t}$ non-uniform steps of length $k^{n} = t^{n+1}-t^{n}$, $n = 1,\ldots,N_{t}$ and define the BDF-2 weights as
\begin{equation}
\beta_0^n = k^n\frac{1+\omega_n}{1+2\omega_n},\quad
\beta_1^n = \frac{(1+\omega_n)^2}{1+2\omega_n},\quad
\beta_2^n = \frac{\omega_n^2}{1+2\omega_n},
\end{equation}
where $\omega_n=k^n/k^{n-1}$, $n=2,\ldots,N_{t}$. In~\cite{Larsson3} it is shown how the time steps can be chosen in such a way that $\beta_0^n\equiv \beta_0$. Therefore, the coefficient matrix is the same in all time steps and only one matrix factorization is needed.

Applying the BDF-2 scheme to ~\eqref{U_setup} or~\eqref{dRBFFD} we arrive at a fully discrete system of equation that reads as follows
\begin{equation}
\begin{bmatrix}
E_{II}+\beta_0L_{II}\; & \beta_0 L_{IB} \\
\beta_0 B_{IB} & E_{BB}+\beta_0 B_{BB}
\end{bmatrix}
\begin{bmatrix}
\vec{u}_I^{\,n}\\ \vec{u}_B^{\,n}
\end{bmatrix}
=
\begin{bmatrix}
\vec{f}_I^{\,n}\\ \vec{f}_B^{\,n}
\end{bmatrix},
\label{impl:system}
\end{equation}
where $E_{II}$ and $E_{BB}$ are the identity matrices of the appropriate size, and
\begin{align}
  \vec{f}_I^{\,n} &= \beta_1^n\vec{u}_{I}^{\,n-1} - \beta_2^n\vec{u}_{I}^{\,n-2}, \label{RHS}\\
  \vec{f}_{B}^{\,n} &= [f(t^n,\ul{x}_{N_{I} +1}),\ldots,f(t^n,\ul{x}_{N})]^{T}. \nonumber
\end{align}
To solve this system we exploit the iterative GMRES method with incomplete LU factorization as preconditioner.

%%%%%%%%%% --- NUMERICAL RESULTS --- %%%%%%%%%%%%
\section{Numerical Experiments}\label{sec:NumExper}

%Since the solution of the European option pricing problem at time $t=T$ has a discontinuity in the first derivative optimal convergence rates of high order methods should not be expected. If the problem was smooth, we could expect  exponential convergence globally, but in our case $V$ is only two times weakly differentiable, i.e., $V\in W_{p}^{2}(\tilde\Omega)$. Embedding theorems proven in~\cite{Rieger} state
%\begin{equation}
%||V||_{L_{q}(\tilde\Omega)} \leq Ch^{2-d\left(\frac{1}{p}-\frac{1}{q}\right)}|V|_{W_{p}^{2}(\tilde\Omega)},
%\end{equation}
%where $h$ is a measure of the node distance, $C$ is some constant, $1\leq p<\infty$, $1 \leq q \leq \infty$, and $d$ is the dimension. Globally we measure the error in the $L_{2}$-norm, therefore, the convergence  can be at most of the order~2.5 for one-dimensional problems and at most of the order~3 for two-dimensional problems. However, for the one-dimensional case we observe a superconvergence for both solver strategies. 

In this section we perform numerical experiments with the presented methods in order to determine the price of a European call option under the assumption of multiple stochastic factors. The design of the experiments is based on~\cite{BenchopMF}. We evaluate the option price for three predetermined asset spot prices that correspond to out-of-the-money, at-the-money, and in-the-money positions. In some cases for some of the selected models there exist semi-analytical solutions. For example, for the Heston model which corresponds to Set~1 of the parameters under the QLSV model, and for the SABR model with zero-correlation between the asset and volatility, i.e., parameter Set~1. Thus, we obtain these values and use them as our reference solutions to study the convergence properties as well as the computational performance of the two localized RBF methods. We test the convergence on node sets with a slightly different number of nodes between the methods, because Cartesian node layouts are not optimal for \mbox{RBF--PUM~\cite{Fornberg2010}}, and for some node densities we may encounter an inconsistency in the results as we refine. Typically, the difference does not exceed one order of magnitude and may with the same likelihood give better and worse approximations. Thus, the errors that we obtain on some repeatedly refined node sets may not align on a straight line when presented using log-log plots. Therefore, we pick points which give a representative picture of the overall convergence rate of RBF--PUM. The values in Tables \ref{TabRBFFEM}, \ref{TabSABR}, \ref{TabHHW}, and \ref{TabHCIR} are obtained on identical node sets for both RBF--FD and RBF--PUM, since there we do not know the exact reference solutions.

We measure the error in the pointwise $l_{\infty}$-norm 
$$
\|\vec{u}-\vec{u}_{\text{ref}}\|_{\infty},
$$
where $\vec{u}_{\text{ref}}$ is the reference solution and is equal to
\begin{itemize}
\item QLSV, Set 1: $\vec{u}_{\text{ref}} = [0.009085, 0.090467, 0.285148]$,
\item SABR, Set 1: $\vec{u}_{\text{ref}} = [0.009545, 0.080717, 0.264368]$.
\end{itemize}
For the plots demonstrating the computational performance, we executed MATLAB implementations of the discussed methods on a laptop equipped with a $2.3$ GHz Intel Core i7 CPU and 16 GB of RAM. Furthermore, for all the comparative figures we use the serial versions of our codes, while the parallelization potentials are presented in the last part of this section.

Additionally, we report tables with option values for the problems where the reference solutions are not readily available.

As we mention in Section~\ref{sec:Methods}, the infinite domain $\Omega$ has to be truncated in order to perform the numerical simulations. We use the following truncated domains
$$
\hat\Omega:=[s_{\text{min}}, s_{\text{max}}]\times[v_{\text{min}}, v_{\text{max}}] = [0, 2]\times[0.001, 1]
$$
for the models with two stochastic factors, 
$$
\hat\Omega := [s_{\text{min}}, s_{\text{max}}]\times[v_{\text{min}}, v_{\text{max}}]\times[r_{\text{min}}, r_{\text{max}}] = [0, 4]\times[0.005, 2]\times[-1, 1]
$$
for the Heston--Hull--White model, and 
$$
\hat\Omega := [s_{\text{min}}, s_{\text{max}}]\times[v_{\text{min}}, v_{\text{max}}]\times[r_{\text{min}}, r_{\text{max}}] = [0, 4]\times[0.005, 2]\times[0, 2]
$$
for the Heston--Cox--Ingersoll--Ross model. We avoid using zero values for $v_{\text{min}}$ because this leads to a singularity in the PDE formulation.

We select the following boundary conditions at the truncated boundaries:
\begin{align}
\label{BC1}& u|_{s=s_{\text{min}}} = 0, \qquad \frac{\p^2 u}{\p s^2}\Big|_{s=s_{\text{max}}} = 0 \,\,\, \text{or} \,\,\, u|_{s=s_{\text{max}}} = s-Ke^{-rt}, \\
& \frac{\p u}{\p v}\Big|_{v=v_{\text{min}}} = \frac{\p u}{\p v}\Big|_{v=v_{\text{max}}} = 0, \\
& \frac{\p u}{\p r}\Big|_{r=r_{\text{min}}} = \frac{\p u}{\p r}\Big|_{r=r_{\text{max}}} = 0.
\end{align}
However, we do not necessarily assign these conditions directly on the solution at the boundary. In case of RBF--PUM, we approximate the boundary conditions by enforcing them inside the differential operator and deriving a reduced form boundary operator $\mathcal{B}$, which is applied at the boundary nodes, while for RBF--FD we assign only the Dirichlet boundary conditions directly at the nodes and for the other boundary nodes we construct the weights in the same way as for the nodes in the inner parts of the computational domain. Our experience shows that this approach works better and provides a more numerically stable localized RBF approximation. We indicate two types of boundary conditions for the far-field boundary in the asset direction. The vanishing second derivative is a correct choice, but for some sets of parameter values we might encounter numerical issues with this approach for the Heston--Hull--White model due to the possibility of the interest rate being negative. Such a situation is challenging for RBF methods, since, e.g., for RBF--PUM no upwind scheme can be constructed. Therefore, as an alternative we suggest the use of a Dirichlet boundary condition, which is asymptotically correct for the Black--Scholes model but biased for the considered models. However, at infinity this bias vanishes. Thus, to diminish the impact of the biased Dirichlet boundary condition the point of truncation $s_{\text{max}}$ must be moved farther away from the origin. 

Although, our methods are featured with freedom of node placement, to discretize the computational domain, we use Cartesian nodes that are uniformly distributed in all directions. The number of nodes is defined as $N_s\times N_v$ for the two-factor problems and $N_s\times N_v \times N_r$ for the three factors models, where $N_s, N_v$, and $N_r$ denote the number of nodes in the asset, volatility, and interest rate directions, respectively. We assign the following values to $N_s, N_v$, and $N_r$
$$
N_s = 2N_v = 2N_r.
$$

%We start the presentation of our results by showing an example of the computed solutions is Figure \ref{sols}. This figure illustrates the importance of studying different pricing models by showing how the value of the option varies when the volatilities are modeled in different ways.
%
%\begin{figure}[H]
%    \centering
%    \begin{subfigure}[t]{0.49\textwidth}
%    \vskip 0pt
%        \includegraphics[height=0.37\textheight]{HSTsol.png}
%    \end{subfigure}
%    \begin{subfigure}[t]{0.49\textwidth}
%    \vskip 0pt
%        \includegraphics[height=0.37\textheight]{SABRsol.png}
%    \end{subfigure}
% \caption{The computed solutions of the pricing problems. \emph{Left:} QLSV model with $\alpha = 0$, $\beta = 1$, and $\gamma = 0$ (Heston).  \emph{Right:} SABR model with $\rho = 0$.}\label{sols}
%\end{figure}

In the following parts of this section we present and discuss the results obtained by the localized RBF methods for each of the problems defined in Section~\ref{sec:Models}. The parameters of our methods used to solve the problems can be found in Appendix \ref{appMethods}.

%Our experiments reveal that RBF--PUM experiences higher sensitivity with respect to the node placement. Therefore, the $N_s$ points shown in the convergence plots for RBF--PUM are carefully chosen to highlight the convergence of the method.

\subsection{The Quadratic Local Stochastic Volatility Model}
\begin{figure}[H]
    \centering
    \begin{subfigure}[t]{0.49\textwidth}
    \vskip 0pt
        \includegraphics[height=0.35\textheight]{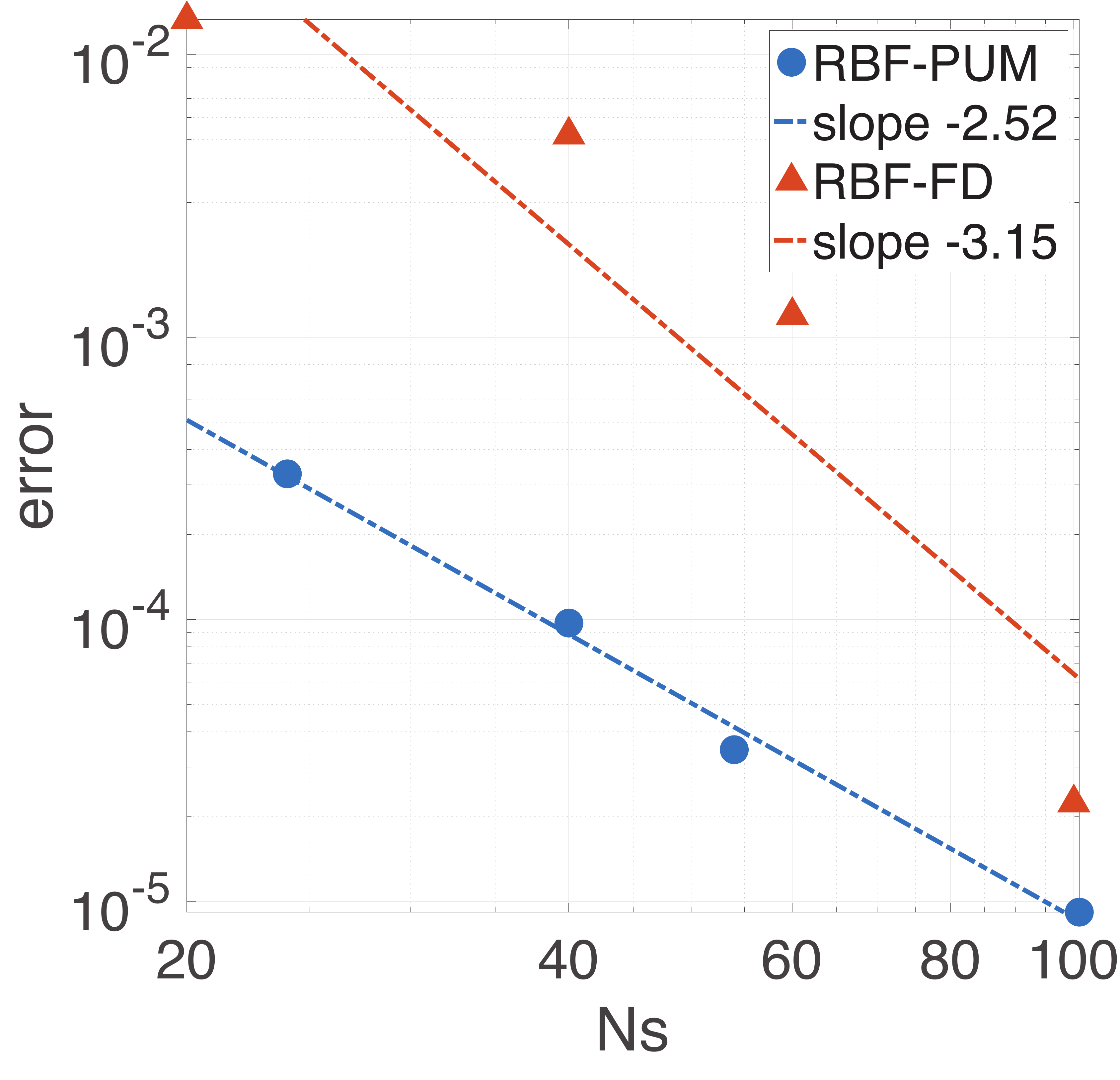}
    \end{subfigure}
    \begin{subfigure}[t]{0.49\textwidth}
    \vskip 0pt
        \includegraphics[height=0.35\textheight]{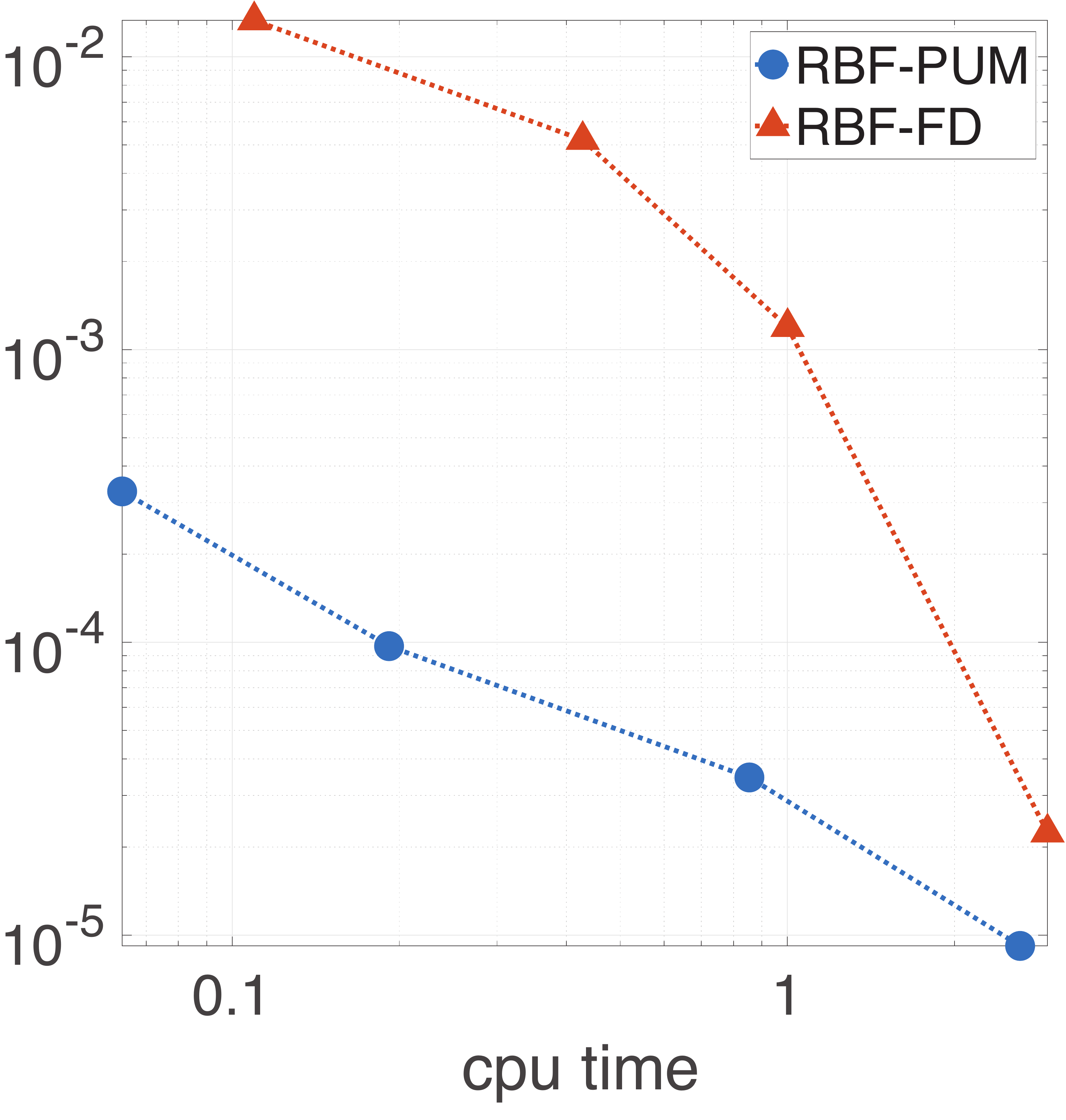}
    \end{subfigure} \caption{Numerical results for the Heston problem. \emph{Left:} Convergence of the localized RBF methods. \emph{Right:} Computational performance of the localized RBF methods measured in seconds of the execution time.}\label{ConvHeston}
\end{figure}

In Figure \ref{ConvHeston} we present the convergence and the computational performance of the two methods for the Heston problem (i.e., QLSV with \mbox{$\alpha=0, \beta=1, \gamma=0$}). 
%Nevertheless, by looking at the orders of convergence it seems as if RBF--FD could potentially reach the accuracy of RBF--PUM at the finer node densities.
%

%\begin{figure}[H]
%%\begin{center}
%\subfigure[Convergence]{\includegraphics[height=0.35\textheight]{HSTconv.eps}}
%\hspace{0.25cm}
%\subfigure[Computational performance]{\includegraphics[height=0.35\textheight]{HSTtime.eps}}
%\caption{Numerical results of the localized RBF methods in terms of convergence and computational performance measured in seconds of the execution time for the Heston problem.}
%\label{ConvHeston}
%%\end{center}
%\end{figure}

It can be seen that both of the methods converge with similar orders, although RBF--PUM performs more efficiently than RBF--FD on the chosen interval. Moreover, both of the methods consume similar computational time for the same number of the discretization nodes, but the accuracy of RBF--PUM dominates. Nevertheless, RBF--FD has a potential to reach the accuracy of RBF--PUM at finer node sets due to a higher order of convergence.

%
%\begin{table}[H] 
%\begin{center}
%\caption{The values of the European call option under the Heston model.}
%\label{TabRBFFEM}
%\begin{tabular}{r c c c | r c c c }
%\hline\hline 
%  \multicolumn{4}{ c |}{RBF--PUM} & \multicolumn{4}{ c }{RBF--FD}  \\   \hline\hline 
% $N_x$  & $S_0=0.75$ & $S_0=1.00$ & $S_0=1.25$ &  $N_x$ &$S_0=0.75$ &  $S_0=1.00$ & $S_0=1.25$ \\  \hline
% $24$  & $0.009060$  & $0.090129$ & $0.285243$  & $20$ &  $0.014568$ & $0.077142$ &  $0.286465$  \\
% $40$  & $0.009182$  & $0.090403$ & $0.285228$  & $40$ &  $0.010303$& $0.085264$ &  $0.285809$ \\
% $54$  & $0.009120$  & $0.090481$ & $0.285121$  & $60$ &  $0.009429$ & $0.089271$ &  $0.285351$ \\
% $101$  & $0.009093$  & $0.090475$ & $0.285139$  & $100$ &  $0.009107$ & $0.090453$ &  $0.285141$ \\
%\hline\hline
%\end{tabular}
%\end{center}
%\end{table}
%

Since we do not have analytical solutions of the QLSV problem with $\alpha=2, \beta=0$, $\gamma=0$, we present the option values for different node set resolutions in Table \ref{TabRBFFEM}. Based on these results we see that both of the methods manage to maintain three digits after the decimal point, although there is a slight difference in the obtained values for at-the-money and in-the-money prices between the methods.
% which could be a consequence of how the methods react on the boundary condition artifacts. 
 The maximum absolute difference between the methods on the finest node set is below $6\cdot10^{-4}$.

\begin{table}[H] 
\begin{center}
\caption{The values of the European call option under the QLSV model with \mbox{$\alpha=2, \beta=0, \gamma=0$}.}
\label{TabRBFFEM}
\begin{tabular}{r c c c | r c c c }
\hline\hline 
  \multicolumn{4}{ c |}{RBF--PUM} & \multicolumn{4}{ c }{RBF--FD}  \\   \hline\hline 
 $N_s$  & $S_0=0.75$ & $S_0=1.00$ & $S_0=1.25$ &  $N_s$ &$S_0=0.75$ &  $S_0=1.00$ & $S_0=1.25$ \\  \hline
 $20$  & $0.005851$  & $0.089819$ & $0.291363$  & $20$ &  $0.007359$ & $0.087353$ &  $0.291477$  \\
 $40$  & $0.005434$  & $0.089478$ & $ 0.291170$  & $40$ &  $0.005493$& $0.088851$ &  $0.290970$ \\
 $60$  & $0.005407$  & $0.089465$ & $0.291152$  & $60$ &  $0.005324$ & $0.088929$ &  $0.290882$ \\
 $100$  & $0.005258$  & $0.089453$ & $0.291114$  & $100$ &  $0.005282$ & $0.088922$ &  $0.290836$ \\
\hline\hline
\end{tabular}
\end{center}
\end{table}

\subsection{The SABR Model}
\begin{figure}[H]
    \centering
    \begin{subfigure}[t]{0.49\textwidth}
    \vskip 0pt
        \includegraphics[height=0.35\textheight]{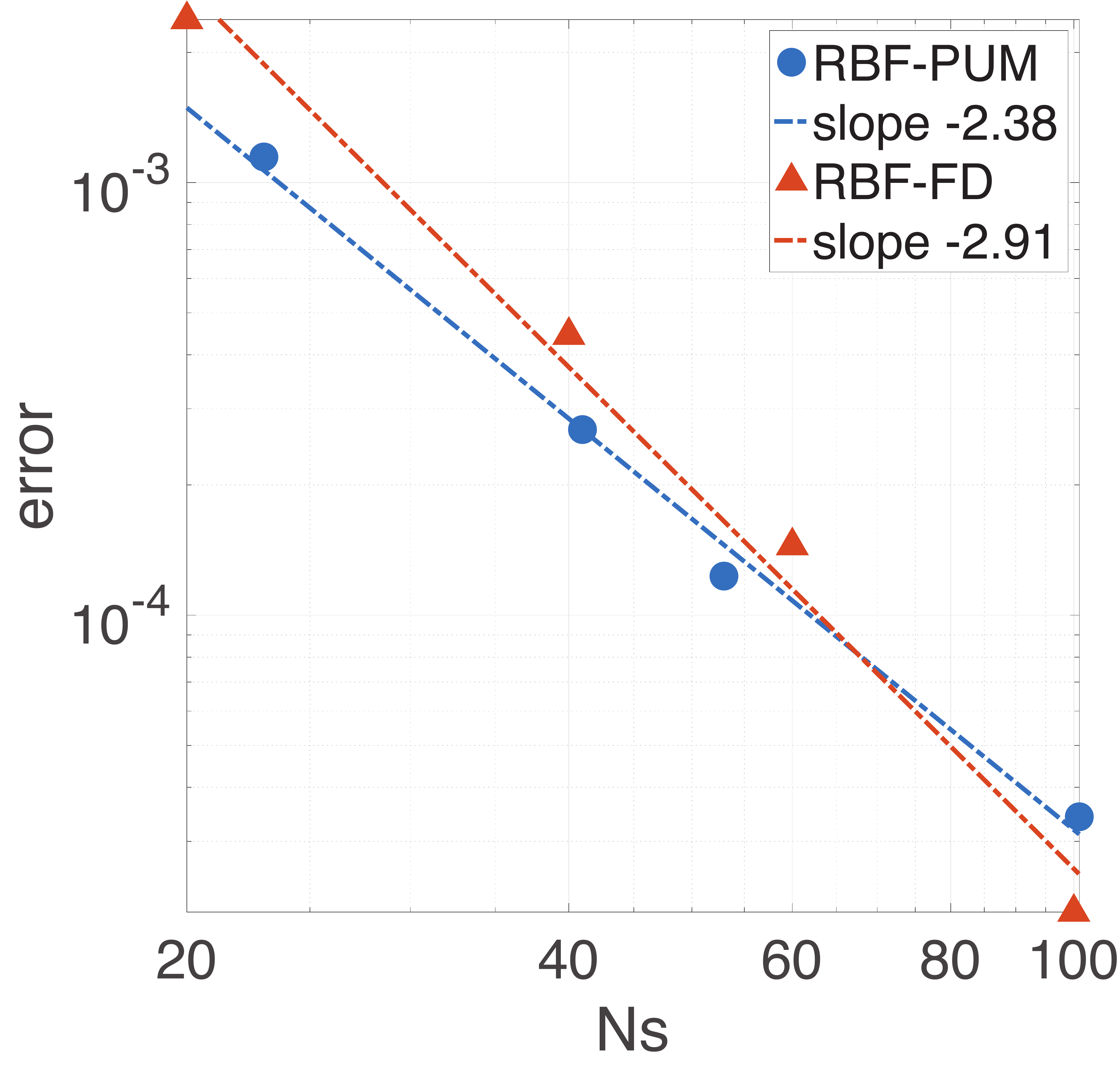}
    \end{subfigure}
    \begin{subfigure}[t]{0.49\textwidth}
    \vskip 0pt
        \includegraphics[height=0.35\textheight]{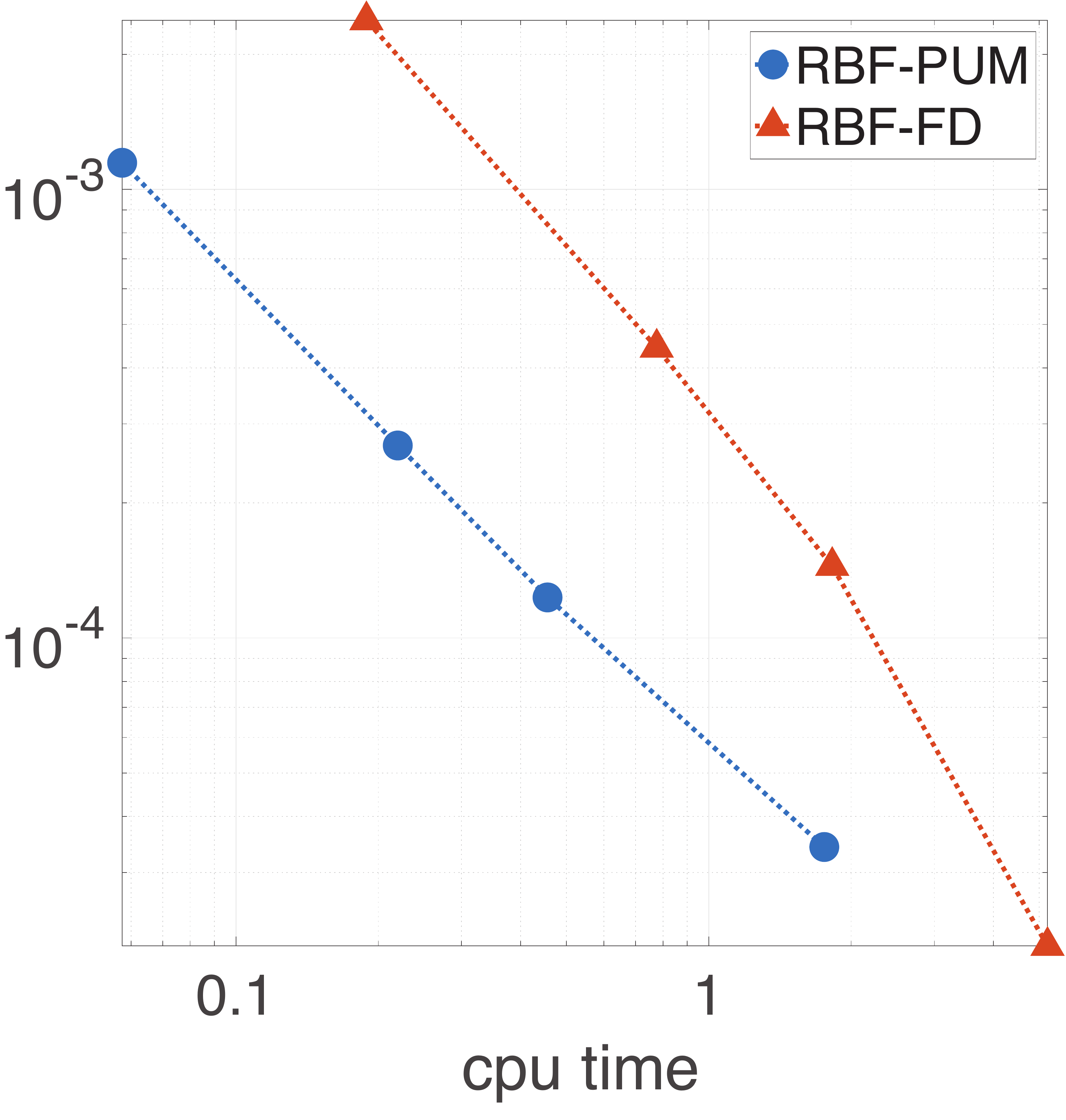}
    \end{subfigure}
 \caption{Numerical results for the SABR problem with $\rho = 0$. \emph{Left:} Convergence of the localized RBF methods. \emph{Right:} Computational performance of the localized RBF methods measured in seconds of the execution time.}\label{ConvSABR}
\end{figure}

In Figure \ref{ConvSABR} we show the convergence and the computational performance of our methods for the SABR problem with $\rho = 0$. 
%\begin{figure}[H]
%%\begin{center}
%\subfigure[Convergence]{\includegraphics[height=0.35\textheight]{SABRconv.eps}}
%\hspace{0.25cm}
%\subfigure[Computational performance]{\includegraphics[height=0.35\textheight]{SABRtime.eps}}
%\caption{Numerical results of the localized RBF methods in terms of convergence and computational performance measured in seconds of the execution time for the SABR problem with $\rho = 0$.}
%\label{ConvSABR}
%%\end{center}
%\end{figure}
Again, we observe that both methods converge with a similar order and demonstrate a more similar accuracy.
%In this case, RBF--PUM starts with a small advantage in accuracy, but RBF--FD takes over towards the finer node densities. 
Nevertheless, RBF--PUM is significantly faster than RBF--FD when it comes to reaching a particular error level. Although, this gap in efficiency may vanish after the algorithm parallelization, because the main overhead comes from the matrix assembly, which is more time-consuming for RBF--FD.

Since we do not have available reference values for the SABR problem with \mbox{$\rho = -0.5$,} we present the option values for different node set resolutions in Table~\ref{TabSABR}. The results shows that both methods manage to converge to the same solutions within a $2\cdot10^{-4}$ maximum absolute difference for the finest node set.

\begin{table}[H] 
\begin{center}
\caption{The values of the European call option under the SABR model with the correlation $\rho=-0.5$.}
\label{TabSABR}
\begin{tabular}{r c c c | r c c c }
\hline\hline 
  \multicolumn{4}{ c |}{RBF--PUM} & \multicolumn{4}{ c }{RBF--FD}  \\   \hline\hline 
 $N_s$  & $S_0=0.75$ & $S_0=1.00$ & $S_0=1.25$ &  $N_s$ &$S_0=0.75$ &  $S_0=1.00$ & $S_0=1.25$ \\  \hline
 $20$  & $0.007379$  & $0.083762$ & $0.269809$  & $20$ &  $0.007028$ & $0.082064$ &  $0.269984$  \\
 $40$  & $0.006017$  & $0.080824$ & $0.268928$  & $40$ &  $0.005641$& $0.080085$ &  $0.268992$ \\
 $60$  & $0.005573$  & $0.080305$ & $0.268649$  & $60$ &  $0.005318$ & $0.080049$ &  $0.268442$ \\
 $100$  & $0.005412$  & $0.080054$ & $0.268513$  & $100$ &  $0.005319$ & $0.079928$ &  $0.268478$ \\
\hline\hline
\end{tabular}
\end{center}
\end{table}
%

%

%

%
%\begin{table}[H] 
%\begin{center}
%\caption{The option values of the European call option under the SABR model.}
%\label{TabRBFFEM}
%\begin{tabular}{r c c c | r c c c }
%\hline\hline 
%  \multicolumn{4}{ c |}{RBF--PUM} & \multicolumn{4}{ c }{RBF--FD}  \\   \hline\hline 
% $N_x$  & $S_0=0.75$ & $S_0=1.00$ & $S_0=1.25$ &  $N_x$ &$S_0=0.75$ &  $S_0=1.00$ & $S_0=1.25$ \\  \hline
% $23$  & $0.010659$  & $0.079571$ & $0.265140$  & $20$ &  $0.010707$ & $0.083098$ &  $0.266444$  \\
% $41$  & $0.009366$  & $0.080447$ & $0.264123$  & $40$ &  $0.009838$& $0.080879$ &  $0.264812$ \\
% $53$  & $0.009639$  & $0.080594$ & $0.264273$  & $60$ &  $0.009521$ & $0.080861$ &  $0.264353$ \\
% $101$  & $0.009579$  & $0.080688$ & $0.264386$  & $100$ &  $0.009544$ & $0.080738$ &  $0.264377$ \\
%\hline\hline
%\end{tabular}
%\end{center}
%\end{table}
%

\subsection{The Heston--Hull--White Model}
The results for this most numerically challenging problem from the list are shown in Table \ref{TabHHW}. The table shows that both methods manage to converge to the same solutions within a $8\cdot10^{-3}$ maximum absolute difference at the highest node densities. It can be observed that the methods tend to converge to slightly different values, which is potentially caused by the numerical issues associated with the negative interest rate values and boundary conditions \eqref{BC1}. The computed option values for the out-of-the-money spot price are significantly different, since that value is much smaller compared to the others.

% reaction of the methods to the approximation of the boundary conditions (\ref{BC1}) which we present in the beginning of this section.

%
\begin{table}[H] 
\begin{center}
\caption{The values of the European call option under the HHW model.}
\label{TabHHW}
\begin{tabular}{r c c c | r c c c }
\hline\hline 
  \multicolumn{4}{ c |}{RBF--PUM} & \multicolumn{4}{ c }{RBF--FD}  \\   \hline\hline 
 $N_s$  & $S_0=0.75$ & $S_0=1.00$ & $S_0=1.25$ &  $N_s$ &$S_0=0.75$ &  $S_0=1.00$ & $S_0=1.25$ \\  \hline
 $20$  & $0.005332$  & $0.109525$ & $0.346429$  & $20$ &  $0.017820$ & $0.128679$ &  $0.357461$  \\
 $30$  & $0.006946$  & $0.117208$ & $0.350123$  & $30$ &  $0.009746$&  $0.123169$ &  $0.356759$ \\
 $40$  & $0.008453$  & $0.125490$ & $0.354327$  & $40$ &  $0.005824$ & $0.122055$ &  $0.359996$ \\
 $50$  & $0.008590$  & $0.127896$ & $0.354546$  & $50$ &  $0.003351$ & $0.120412$ &  $0.360258$ \\
\hline\hline
\end{tabular}
\end{center}
\end{table}

\subsection{The Heston--Cox--Ingersoll--Ross Model}
Finally, we present the results for the HCIR problem in Table \ref{TabHCIR}. The option values are similar to the values obtained for the HHW problem, which verifies the trustworthiness of our solvers, since they are evaluated on the same parameter sets, and the only difference between these problems is in the chosen stochastic model for the dynamics of the interest rate $R_t$. The table shows that both methods manage to converge to the same solutions again within a $8\cdot10^{-3}$ maximum absolute difference at the highest node density. Here, again we see a slight difference in the converged values. As it can be seen in the previously presented experiment, there is a significant relative difference between the option values out-of-the-money, since that value is small. 

\begin{table}[H] 
\begin{center}
\caption{The values of the European call option under the HCIR model.}
\label{TabHCIR}
\begin{tabular}{r c c c | r c c c }
\hline\hline 
  \multicolumn{4}{ c |}{RBF--PUM} & \multicolumn{4}{ c }{RBF--FD}  \\   \hline\hline 
 $N_s$  & $S_0=0.75$ & $S_0=1.00$ & $S_0=1.25$ &  $N_s$ &$S_0=0.75$ &  $S_0=1.00$ & $S_0=1.25$ \\  \hline
 $20$  & $0.003478$  & $0.108315$ & $0.346736$  & $20$ &  $0.019821$ & $0.130269$ &  $0.353338$  \\
 $30$  & $0.004580$  & $0.117966$ & $0.351828$  & $30$ &  $0.009327$& $0.122469$ &  $0.357499$ \\
 $40$  & $0.005405$  & $0.119717$ & $0.351985$  & $40$ &  $0.004659$ & $0.120718$ &  $0.360338$ \\
 $50$  & $0.006225$  & $0.122347$ & $0.352823$  & $50$ &  $0.002748$ & $0.120701$ &  $0.360891$ \\
\hline\hline
\end{tabular}
\end{center}
\end{table}

\subsection{Parallelization}
\begin{figure}[H]
    \centering
    \begin{subfigure}[t]{0.49\textwidth}
    \vskip 0pt
        \includegraphics[height=0.35\textheight]{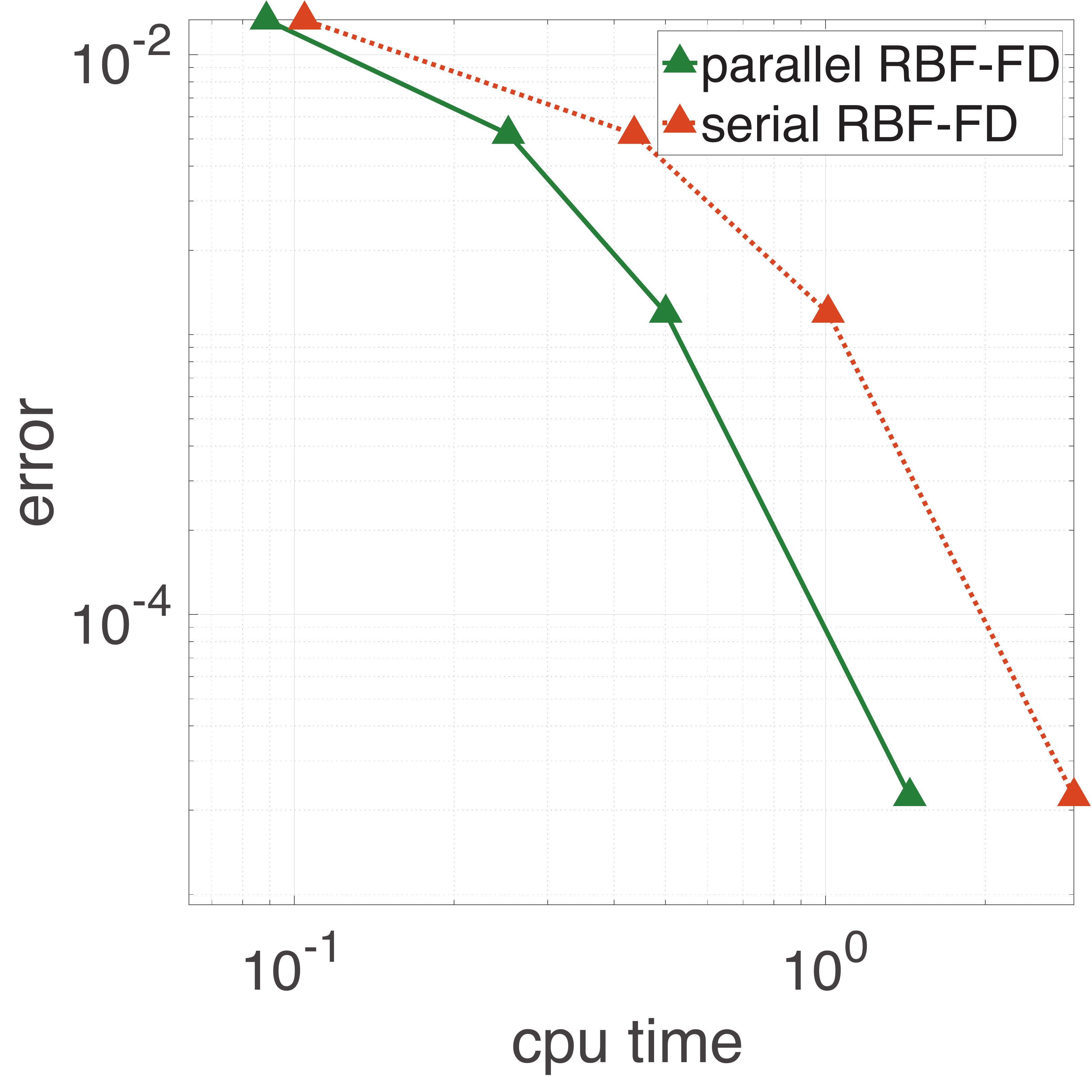}
    \end{subfigure}
    \begin{subfigure}[t]{0.49\textwidth}
    \vskip 0pt
        \includegraphics[height=0.35\textheight]{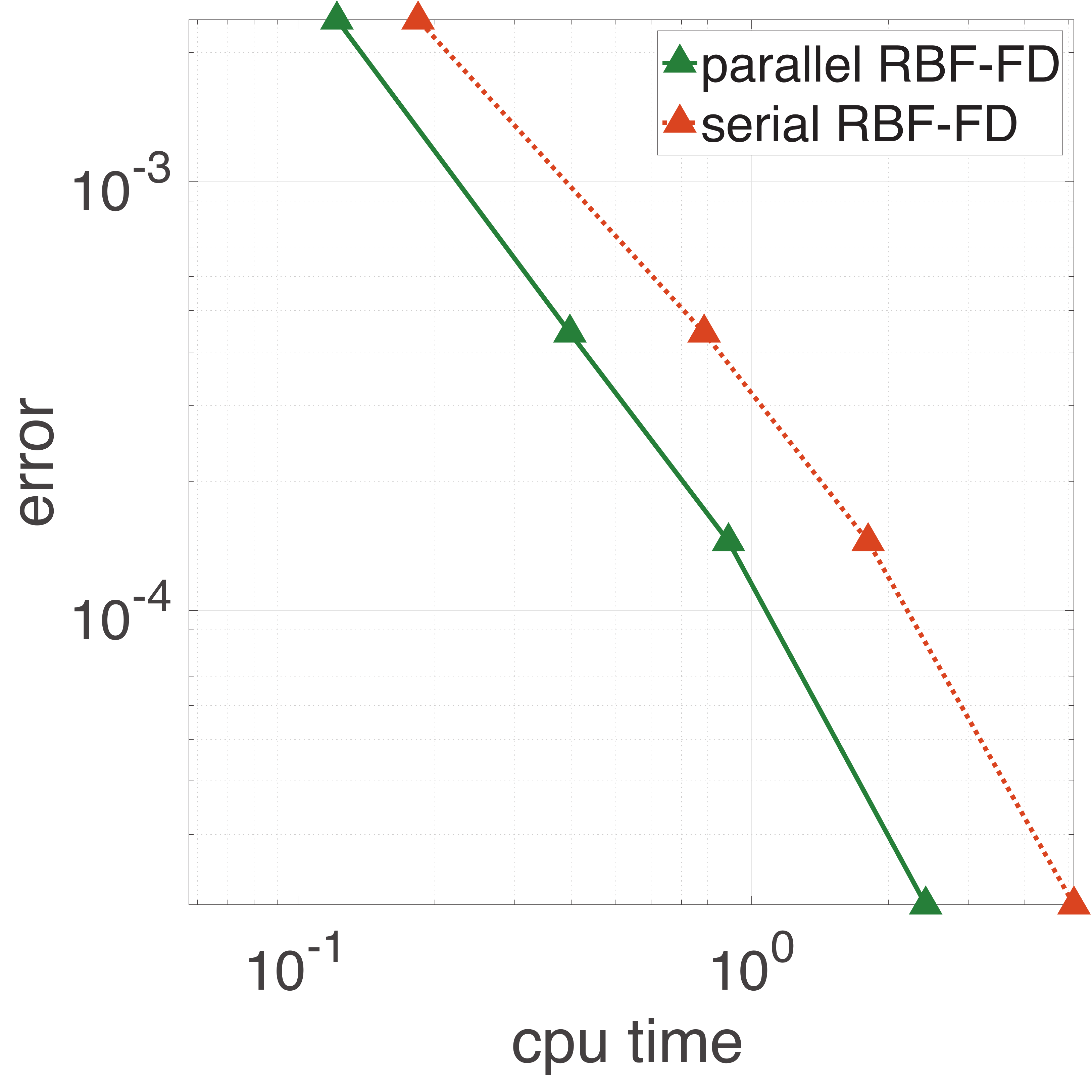}
    \end{subfigure}
 \caption{Computational performance of parallel and serial RBF--FD solvers measured in seconds of the execution time. \emph{Left:} The Heston problem. \emph{Right:} The SABR problem.}\label{parallel}
\end{figure}

As mentioned above, both of the methods have parallelization potentials. We pick RBF--FD to demonstrate the improvements in the computational performance when the calculation of the differentiation weights is parallelized. For this purpose we use the \texttt{parfor} functionality available in the Parallel Computing Toolbox by MATLAB, which scatters the computational load across $n_w$ available parallel workers. On our machine we had $n_w=4$. In Figure~\ref{parallel} we present the computational performance of the parallel implementations compared with the computational performance of the serial codes for the Heston and SABR problems. The red dotted line is the same one seen in the right-hand side panels of Figures \ref{ConvHeston} and \ref{ConvSABR}.

%\begin{figure}[H]
%%\begin{center}
%\subfigure[The Heston problem]{\includegraphics[height=0.35\textheight]{HSTparallel.eps}}
%\hspace{0.25cm}
%\subfigure[The SABR problem]{\includegraphics[height=0.35\textheight]{SABRparallel.eps}}
%\caption{Computational performance of parallel and serial RBF--FD solvers measured in seconds of the execution time.}
%\label{parallel}
%%\end{center}
%\end{figure}

RBF--PUM can be parallelized in a similar way as RBF--FD by running in parallel the operations inside the patches.

The results indicate that by parallelization the execution time of RBF--FD can be decreased by several times. It shows a decrease from $5.13$ to $2.41$ seconds of run time for the SABR problem on the finest node set. Moreover, the figures suggest that the computational savings increase as a more accurate solution is required. 

Finally, these speedups may be much greater if the computational load is scattered on a high performance computing cluster that has more than $4$ employable parallel workers. Nevertheless, we need to stress that the parallel RBF--FD solver still contains a non-parallelized time integration part that is based on the BDF-2 method with GMRES for the iterative solution of the linear systems. The operations of matrix-vector multiplications within GMRES can also be parallelized~\cite{tominec, Tillenius,bollig2012solution}, but for that level of performance optimization it would be more suitable to use a compiled programming language and more advanced hardware, which is out of the scope of this study.

%%%%%%%%%% --- CONCLUSIONS --- %%%%%%%%%%%%
\section{Conclusions}\label{sec:Conclusions}
In this paper we study the usage of localized RBF methods, namely RBF--PUM and \mbox{RBF--FD}, for numerical pricing of financial derivatives under models with multiple stochastic factors. 

We choose advanced pricing models that are of interest for the financial industry. In these models the volatilities, interest rates, or both, are modeled as stochastic processes, in addition to the stochastic asset dynamics, therefore spawning challenging time-dependent partial differential equations in two or three spatial dimensions as mathematical problems that need to be solved (in most cases possible only numerically), in order to obtain the option prices. As an appropriate approach to solving these problems, we demonstrate in detail the best properties of the two localized RBF methods, such as sparsity and high convergence orders, and show how numerical solvers for the posed problems can be designed and implemented. Moreover, we test and compare our solvers in a fair setting. We demonstrate the convergence properties of the numerical methods and their computational performance. Furthermore, we discuss the parallelization potentials and gains of the presented methods. 

The results demonstrate the capability of both methods to solve the posed problems to an adequate accuracy in reasonable time (a few seconds). Both methods show similar orders of convergence, which can be further optimized by more elaborate choices of the parameters, such as the sizes of patch overlaps for RBF--PUM and the stencil sizes for RBF--FD. In our tests, RBF--FD shows slightly higher orders of convergence and is able to achieve the accuracy of RBF--PUM at the finer node layouts. Moreover, in the case of serial algorithms, RBF--PUM appears to be faster than RBF--FD in some cases, but after parallelization this gap vanishes due to the distribution of the task of weights computation over multiple parallel workers.  

Overall, both methods possess very similar performance features and it is difficult to advice on which method should be selected for a particular application. Nevertheless, from the experiments it is clear that both methods are suitable for problems with multiple factors and have high potential to tackle high-dimensional PDEs that arise from the problems of mathematical finance. 

% All said, both methods seem to possess very similar performance features and it is hard to know in advance which one would perform better when it comes to a particular practical usage. Furthermore, as our methods performed well for the QLSV and SABR problems, the results for HHW and HCIR indicated how important it is to accurately enforce the proper boundary conditions. Due to the instabilities that our methods encounter when strongly imposing non-Dirichlet boundary conditions, our oversimplified approximation of the boundary conditions appeared as a bias in the computed solutions for those stochastic interest rate problems.

%Nevertheless, overall conclusion is that the presented methods possess high potential to tackle advanced high-dimensional PDEs in finance and that with some further development they could find a place between the standard grid based methods for low dimensional problems and the Monte Carlo simulations for high dimensional problems. 

%We present numerical results in 1D, 2D, and 3D that demonstrate the good
%properties of our method. Due to the sparsity of the discretization matrix and the possibility of local node refinement, we believe that our suggested method has the potential to work well also in higher dimensions.

\section*{\large{Acknowledgement}}

{\footnotesize{The authors would like to thank Elisabeth Larsson and Lina von Sydow for fruitful discussions and for proofreading the manuscript.}}

\section*{\large{Declaration of Interest}}
{\footnotesize{The authors report no conflicts of interest. The authors alone are responsible for the content and writing of the paper.}}

%%% REFERENCES %%%
%\newpage
\bibliographystyle{unsrt}
{\footnotesize
\bibliography{bibpaper}}
\appendix

\section{Method Parameters}
\label{appMethods}
Below are the values of the various method parameters that were used to obtain the presented results.
\begin{itemize}

\item Values for GMRES: We use the \texttt{nofill} setting for the incomplete LU factorization to produce the preconditioner for the iterative solver. We set the tolerance to $10^{-8}$ for all our experiments. Once the residual drops below the selected tolerance, we acknowledge the iteration as converged. To speed up the convergence we use the values from the previous time step as the initial value for the next iteration.

\item Values for RBF-PUM: We use multiquadric RBFs for all the experiments. The shape parameter $\varepsilon$ is initially chosen as 
$\varepsilon = 0.17/h - 0.8$ and then slightly adjusted for each model individually. Here \mbox{$h=(s_{\max}-s_{\min})/(N_s-1)$.} The number of patches is chosen in such a way to have approximately 130 computational nodes in the interior patches. The radius of the patches is computed as $\sqrt{2}H(1+\delta)$, where $H$ is the half spacing distance between the patch centers and $\delta=0.2$. The number of partitions along the spacial dimensions is chosen according to the ratio $P_s=2P_v=2P_r$.

\item Values for RBF-FD: For the approximations of weights in the QLSV and SABR problems, we use PHSs of degree $q = 5$ augmented with monomials of degree  $q_p=5$ which span a polynomial space of size $m_j = 21$. The stencil sizes are chosen to be $n_j := 3\cdot m_j = 63$. As the HHW and HCIR problems require more stability, the degree of monomials and PHS are decreased to $q=3$ and  $q_p=3$, which gives $m_j=20$ and we chose $n_j := 5\cdot m_j = 100$.

\end{itemize}

\section{Algorithm for RBF-PUM}
\label{A:rbfpum}

\begin{algorithm}[H]
\caption{Valuing options by RBF-PUM}
\label{AlgPEN}
\begin{algorithmic}[1]
  \State Define a set of computational nodes $\bold{x}$.
  \State Define a domain partitioning $\{\Omega_j\}_{j=1}^{P}$.
  \State Construct the partition of unity weights $w_j$ using \eqref{PUweight}.
  \State Construct the matrix as in \eqref{impl:system} by running the following \textbf{for} loop.
  \For{each patch from 1 to $P$}:
    \State Compute distances between local nodes in a patch.
    \State Compute local differentiation matrices.
    \State Construct a local differential operator.
    \State Insert the local differential operator into the global differential operator in the position specified by the node set indexing.
  \EndFor
  \State Initialise $\vec{u}^{\,0}= \text{Payoff}$.
  \State Compute incomplete LU factors, iL and iU, for the matrix in \eqref{impl:system}.
  \State Define the tolerance tol for GMRES.
  \For{each time step from 2 to $N_t$}:
    \State Define rhs according to \eqref{RHS}.
    \State Solve for $\vec{u}^{\, n}$ using GMRES $\vec{u}^{\, n}=\text{gmres}(\text{rhs},\text{iL},\text{iU},\vec{u}^{\, n-1},\text{tol})$.
    \State Update $\vec{u}^{\, n-2}=\vec{u}^{\, n-1}$ and $\vec{u}^{\, n-1}=\vec{u}^{\, n}$.
  \EndFor
\end{algorithmic}
\end{algorithm}

\newpage
\section{Algorithm for RBF-FD}
\label{A:rbffd}

\begin{algorithm}[H]
\caption{Valuing options by RBF-FD}
\label{AlgRBFFD}
\begin{algorithmic}[1]
  \State Define a set of computational nodes $\bold{x}$.
  \State Choose the polynomial degree $q_p$ and the stencil sizes $n_j$.
  \State Construct the matrix as in \eqref{impl:system} by running the following \textbf{for} loop.
  \For{each node from 1 to $N$}:
    \State Estimate its $n_j-1$ nearest neighbors using $k$-d tree algorithm. 
    \State Compute local distance matrix.
    \State Compute local RBF and monomial matrices as in (\ref{eq:D2}).
    \State Compute the right-hand side vector as in (\ref{eq:D2}).
    \State Solve the linear system (\ref{eq:D2}) to obtain the weights for that node.
    \State Store the weights in the global differential operator at the position specified by the node set indexing.
  \EndFor
  \State Initialise $\vec{u}^{\,0}= \text{Payoff}$
  \State Compute incomplete LU factors, iL and iU, for the matrix in \eqref{impl:system}.
  \State Define the tolerance tol for GMRES.
  \For{each time step from 2 to $N_t$}:
    \State Define rhs according to \eqref{RHS}.
    \State Solve for $\vec{u}^{\, n}$ using GMRES $\vec{u}^{\, n}=\text{gmres}(\text{rhs},\text{iL},\text{iU},\vec{u}^{\, n-1},\text{tol})$.
    \State Update $\vec{u}^{\, n-2}=\vec{u}^{\, n-1}$ and $\vec{u}^{\, n-1}=\vec{u}^{\, n}$.
  \EndFor
\end{algorithmic}
\end{algorithm}

\end{document}